\documentclass[prb, twocolumn,showpacs,preprintnumbers,amsmath,amssymb, superscriptaddress]{revtex4}

\usepackage{graphicx,color,amsmath,amssymb}

\begin{document}
\newcommand{\kk}{{\bf k}}
\newcommand{\Q}{{\bf Q}}
\newcommand{\q}{{\bf q}}
\newcommand{\gk}{g_\textbf{k}}
\newcommand{\ee}{\tilde{\epsilon}^{(1)}_\textbf{k}}
\newcommand{\HH}{\mathcal{H}}

\title{Superconductivity and local non-centrosymmetricity in crystal lattices}
\author{Mark H. Fischer}
\affiliation{%
Department of Physics, Cornell University, Ithaca, New York 14853, USA}
\affiliation{%
Institut f\"ur Theoretische Physik, ETH Z\"urich, 8093 Z\"urich, Switzerland
}%
\author{Florian Loder}
\affiliation{%
Center for Electronic Correlations and Magnetism, Institute of Physics, Universit\"at Augsburg, D-86135 Augsburg, Germany
}%
\author{Manfred Sigrist}
\affiliation{%
Institut f\"ur Theoretische Physik, ETH Z\"urich, 8093 Z\"urich, Switzerland
}%

\date{\today}

\begin{abstract}
Symmetry of the crystal lattice can be a determining factor for the structure of Cooper pairs in unconventional superconductors. In this study we extend the discussion of superconductivity in non-centrosymmetric materials to the case when inversion symmetry is missing locally, but is present on a global level. Concretely, we investigate the staggered non-centrosymmetricity within a regular sublattice structure, in some analogy to the discussion of superconductivity in antiferromagnetic systems. Three crystal structures are analyzed in detail as illustrative examples for the extended classification of Cooper-pairing channels. One of the cases may be relevant for the class of iron-pnictide superconductors. 
\end{abstract}

\maketitle

\section{Introduction}
\label{sec:intro}
Shortly after the seminal paper by Bardeen, Cooper, and Schrieffer describing superconductivity through pairing of electrons of equal energy and opposite spin and momentum,\cite{bardeen:57} Anderson realized that the existence of such degenerate electron pairs would be guaranteed quite generally by time-reversal symmetry.\cite{anderson:59} Indeed, removing time-reversal symmetry by an external magnetic field, magnetic impurities, or ferromagnetic order substantially weakens or even suppresses superconductivity in the spin-singlet channel. Later, Baltensperger and Str\"assler demonstrated that spin-singlet superconductivity and antiferromagnetism can coexist for an appropriate pair structure.\cite{baltensperger:63} 
In such systems, staggered moments break time-reversal symmetry only on sublattices.
However, time-reversal operation may be undone globally by exchanging the two sublattices. In this case, the spin-singlet pair wavefunction has dominant amplitudes for the two electrons being on different sublattices. 

For Cooper pairing of electrons in the spin-triplet configuration, Anderson showed several years later that an additional discrete symmetry is needed, namely inversion symmetry.\cite{anderson:84} The discovery of superconductivity in crystals lacking an inversion center and yet showing features usually attributed to spin-triplet pairing, therefore, attracted much attention in recent years. Non-centrosymmetricity affects 
the electronic spectrum through symmetry specific antisymmetric spin-orbit coupling. 
Spin-triplet superconductivity is not simply suppressed in favor of spin-singlet pairing, but  
actually electrons pair with a mixed-parity structure combining a spin-singlet and -triplet component. 

Since in the context of time-reversal-symmetry breaking and superconductivity the effect of both,
ferromagnetic and antiferromagnetic order on the Cooper-pair formation have been studied, it is natural to extend the recent discussion of globally non-centrosymmetric superconductivity to its staggered form. 
Recently, Yanase has analyzed the case of locally broken inversion symmetry due to stacking faults, where the global inversion symmetry is retained because of the random distribution of these faults.\cite{yanase:2010} In the present study, we will generalize the discussion of broken inversion symmetry from a ferro-type to the antiferro-type. 

After introducing first a general formulation for the antiferro-type of ``non-centrosymmetric'' lattices, we will discuss two examples in detail to illustrate the influence on superconductivity and then apply the results to a crystal structure as found in the iron pnictide superconductors. 
The underlying crystal symmetry for all three systems is tetragonal and can be characterized by a specific sublattice structure of two distinct types of sites or bonds yielding a doubling of the ordinary unit cell. 
Each of them has a different subgroup of $ D_{4h} $ leaving the sublattice structure invariant. The examples then differ in that the first and third example have a sublattice lacking inversion symmetry, while the second example lacks inversion symmetry only on the bonds connecting the two sublattices.

\section{Single particle Hamiltonian}
\label{sec:singleparticle}
\begin{figure}[t]
    \begin{center}
	\includegraphics{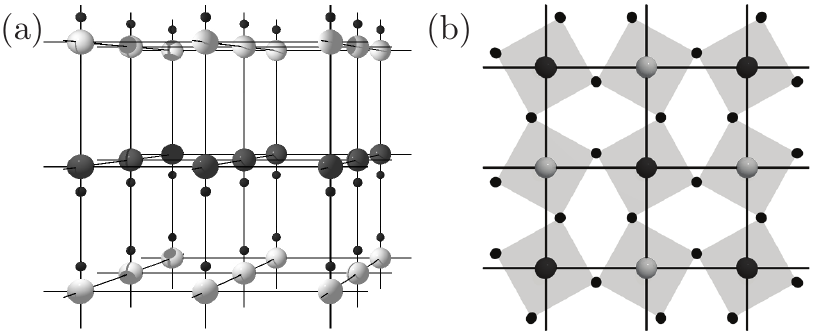}
    \end{center}
    \caption{The two example crystal structures analyzed in detail in Secs.~\ref{sec:layers} and \ref{sec:bonds}. (a) Inversion-symmetry lacking layers that are stacked along the $z$ direction in a staggered way. The symmetry center lies between the layers and the crystal has a symmorphic structure. (b) Top view of the crystal structure with O$_{6}$ octahedra rotated around the $c$ axis leading to a doubling of the unit cell. While both sublattices still retain inversion symmetry, the bonds do not as the rotation shifts the O ions off the bonds. This crystal structure with its symmetry center on one of the sublattices is non-symmorphic.}
    \label{fig:structures}
\end{figure}

Before looking at these specific examples, we introduce a general formalism for superconductivity in a lattice with a non-centrosymmetric sublattice structure. While such a crystal has centers of inversion, the lattice structure includes local violations of inversion symmetry (see Fig.~\ref{fig:structures}), which yield a staggered form of antisymmetric spin-orbit coupling. This can be incorporated into the kinetic energy by defining a folded Brillouin zone with two bands characterized by the wave vector $ \Q $ ($2 \Q $ is a reciprocal lattice vector). Thus, we define the operators,
\begin{equation}
    c_{\alpha \kk s} = \left\{\begin{array}{ll} c_{\kk s} & \alpha = 1\\ c_{\kk + \Q s} & \alpha = 2,\end{array}\right.
    \label{eq:twoband}
\end{equation}
where we use $ \alpha = 1,2 $ as band indices. 

\subsection{Diagonal single-particle Hamiltonian}

First, we consider the general structure of the single-particle Hamiltonian in the two-band language diagonal in the electron operators, i.e. the general form of the kinetic energy.
This part of the Hamiltonian is understood in terms of hopping.

\subsubsection{Spin-independent terms} 

The spin-independent part is given by 
\begin{equation}
    \HH = \sum_{\alpha, \alpha'}\sum_{\kk, s}\Xi_{\kk \alpha\alpha'}c_{\alpha \kk s}^{\dag}c_{\alpha'\kk s}^{\phantom{\dag}},
    \label{eq:sph0}
\end{equation}
where for ordinary hopping the energy term $\Xi_{\kk \alpha\alpha'} = \Xi_{\kk\alpha\alpha} \delta_{\alpha \alpha'} $ is diagonal in the band index. We may explicitly write 
\begin{multline}
    \HH = \sum_{\kk,s}\Big[(\varepsilon_{\kk}^{\rm intra} - \mu + \varepsilon_{\kk}^{\rm inter})c_{1 \kk s}^{\dag}c_{1\kk s}^{\phantom{\dag}}\\ + (\varepsilon_{\kk}^{\rm intra} - \mu - \varepsilon_{\kk}^{\rm inter})c_{2 \kk s}^{\dag}c_{2\kk s}^{\phantom{\dag}}\Big].
    \label{eq:Xi}
\end{multline}
Here, $\varepsilon_{\kk}^{\rm intra} = \varepsilon_{\kk + \Q}^{\rm intra}$ is an intra-sublattice term, i.e. represents hopping between sites of the same sublattice type. 
Then, $\varepsilon_{\kk}^{\rm inter} = -\varepsilon_{\kk + \Q}^{\rm inter}$ is correspondingly an inter-sublattice term due to hopping between sites of different sublattices. 
It is now useful to introduce Pauli matrices $(\tau^0, \vec{\tau})$ for the orbital space with which the matrix element in Eq.~\eqref{eq:Xi} simplifies to 
\begin{equation}
    \Xi_{\kk \alpha\alpha'} = (\varepsilon_{\kk}^{\rm intra} - \mu)\tau^0_{\alpha\alpha'} + \varepsilon_{\kk}^{\rm inter}\tau^{3}_{\alpha\alpha'}.
    \label{eq:Xi1}
\end{equation}
A sublattice dependent chemical potential $\mu_{A (B)} = \mu \pm \Delta\mu$ would lead to an additional spin-independent term in the Hamiltonian of the form
\begin{equation}
    \HH = \sum_{\kk, s}\Delta\mu(c_{1 \kk s}^{\dag}c_{2\kk s}^{\phantom{\dag}}+{\rm h.c.})
    \label{eq:hchem}
\end{equation}
or again in terms of Pauli matrices
\begin{equation}
  \Xi_{\kk\alpha\alpha'} = \Delta\mu\tau^1_{\alpha\alpha'},
    \label{eq:hchem2}
\end{equation}
off-diagonal in the band index. 
Note that time-reversal symmetry leads to the condition that 
\begin{equation}
\Xi_{\kk \alpha \alpha'} = \Xi_{-\kk \alpha \alpha'}
\end{equation}
and $ \Xi_{\kk \alpha \alpha'} $ is spin independent. 

The four Pauli matrices $(\tau^0, \vec{\tau})$ for the band part are easily interpreted
in the sublattice and two-band notion. Matrix elements independent of sublattice are diagonal represented by $\tau^0$ for intra-sublattice and by $ \tau^3 $ for inter-sublattice processes. Analogously, inter-band hybridization is incorporated in $\tau^1$ and $ \tau^2 $ for intra- and inter-sublattice couplings, respectively. This is summarized in Table~\ref{tab:taus}. 
\subsubsection{Spin-dependent terms}

We now turn to the spin-dependent part of the Hamiltonian, 
\begin{equation}
    \HH =  \sum_{\alpha, \alpha'}\sum_{\kk} \sum_{s,s'} \Gamma_{\kk\alpha\alpha'}^{ss'}c_{\alpha \kk s}^{\dag}c_{\alpha'\kk s'}^{\phantom{\dag}}.
    \label{eq:hgen}
\end{equation}
This single-particle Hamiltonian will be written as a tensor product of a spin and a band part. 
In the following, summation over repeated indices is implicit.

First, we consider terms which are based on intra-sublattice contributions, connecting only sites of the same sublattice. These can be written as
\begin{equation}
  \Gamma_{\kk\alpha\alpha'}^{ss'}  = \vec{f}_{\kk}^{\,0} \cdot \vec{\sigma}_{ss'} \otimes \tau^0_{\alpha \alpha'} + \vec{f}_{\kk}^{\,1} \cdot \vec{\sigma}_{ss'} \otimes \tau^1_{\alpha \alpha'} \; ,
\label{eq:Gammaintra}
\end{equation}
involving intra- and inter-band terms, $ \vec{f}_{\kk}^{\,0} $ and  $ \vec{f}_{\kk}^{\,1} $. Analogously the inter-sublattice part is given by
\begin{equation}
  \Gamma_{\kk\alpha\alpha'}^{ss'}  = \vec{g}_{\kk}^{\,2} \cdot \vec{\sigma}_{ss'} \otimes \tau^2_{\alpha \alpha'} + \vec{g}_{\kk}^{\,3} \cdot \vec{\sigma}_{ss'} \otimes \tau^3_{\alpha \alpha'} \; .
\label{eq:Gammainter}
\end{equation}
These terms are important, if time-reversal and inversion symmetry are violated. Time reversal $ \hat{T} $ and inversion $ \hat{I} $ operate as
\begin{equation}
  \hat{T} \vec{f}_{\kk}^{\,a} = - \vec{f}_{-\kk}^{\,a} 
  \label{eq:timereversal}
\end{equation}
and
\begin{equation}
  \hat{I} \vec{f}_{\kk}^{\,a} =  \vec{f}_{-\kk}^{\,a} 
\end{equation}
and analogously for $ \vec{g}_{\kk}^{\,a} $. 

For illustration, let us look at a few generic examples. Zeeman coupling of all spins to a uniform magnetic field $ \vec{H}_0 $ is implemented by $ \vec{f}_{\kk}^{\,0} =g\mu_{\rm B}\vec{H}_0 $ and, correspondingly, a staggered field $ \vec{H}_Q $ (opposite for the electron spins on the two sublattices) is represented as $ \vec{f}_{\kk}^{\,1} = g\mu_{\rm B}\vec{H}_Q $ (analogous to the sublattice dependent chemical potential), both being on-site-coupling (intra-sublattice) terms. According to Eq.~\eqref{eq:timereversal}, they introduce a violation of time-reversal symmetry. Spin-dependent hopping terms connecting the same or different sublattices can be written as
\begin{equation}
  \vec{f}_{\kk}^{\,0} = \vec{\lambda}_{\kk}^{\rm intra} \quad \mbox{and} \quad 
  \vec{g}_{\kk}^{\,3} = \vec{\lambda}_{\kk}^{\rm inter} ,
\label{f0-g3}
\end{equation}
respectively. More important for our subsequent discussion are ''staggered'' spin-orbit coupling terms which correspond to
\begin{equation}
  \vec{f}_{\kk}^{\,1} = \vec{\zeta}_{\kk}^{\rm intra} \quad \mbox{and} \quad 
\vec{g}_{\kk}^{\,2} = \vec{\zeta}_{\kk}^{\rm inter} ,
\label{f1-g2}
\end{equation}
for intra- and inter-sublattice hopping, respectively.

\subsubsection{Symmetry considerations}

We consider now some symmetry aspects, whereby the sublattice structure again plays an important role. 
We introduce $ {\cal G} $ as the generating point group and denote by $ {\cal G}'$ the subgroup of operations respecting the sublattice structure. All other operations in ${\cal G}\setminus {\cal G}'$ interchange the two sublattices. As we consider centrosymmetric crystals, we request that the inversion is contained in $ {\cal G}$. However, inversion may or may not be contained in ${\cal G}'$. In the former case, the symmetry center lies on one of the two sublattices and the operations in ${\cal G}\setminus{\cal G}'$ have to be accompanied with a translation undoing the interchange of the sublattices. This means that the space group of these crystals does not contain ${\cal G}$ as a subgroup and is therefore non-symmorphic. In the latter case, the center of inversion lies between the sublattices. This can lead to both, symmorphic and non-symmorphic crystal structures as we will see in the following. 

As noted above, the diagonal terms of the single-particle Hamiltonian have a tensor product structure, consisting of the momentum-dependent spin part  ($ \epsilon_{\kk} \sigma^0 $, $ \vec{f}_{\kk} \cdot \vec{\sigma} $ and $ \vec{g}_{\kk} \cdot \vec{\sigma} $) and the band part expressed by the $ \tau $-matrices. Therefore, we may classify these terms by means of irreducible representations of $ {\cal G} $, as $ R \otimes R' $. 
The symmetry operations $ g \in {\cal G} $ on the momentum-dependent spin part act as
\begin{equation}
  g \epsilon_{\kk} = \epsilon_{D_g^- \kk}\;\; {\rm and}\;\; g  \vec{f}_{\kk} = D_g^+ \vec{f}_{D_g^- \kk},
\end{equation}
where $ D_g^- $ is the corresponding operation of element $ g$ on a vector and  $ D_g^+ $ on a pseudo-vector. For the band part, 
it is easy to see that $ \tau^0$ and $\tau^3$ do not change under such an interchange of the sublattices, such that they belong to the trivial irreducible representation $ A_{1g} $ 
of $ {\cal G} $. On the other hand, terms with $\tau^1$ and $\tau^2$ change sign under the interchange of sublattices and belong to an irreducible representation $ \Gamma' $ specific to $ {\cal G} $ and the sublattice structure. Depending on whether inversion is an element of ${\cal G}'$ or not, $\Gamma'$ will be an even or odd representation. 

For illustration, we consider two specific examples for a lattice with tetragonal symmetry with $ {\cal G} =D_{4h} $, which will be discussed in more detail below. 
The first example has a sublattice structure such that the $A$- and $B$-sublattices form alternating layers
along the $z$-axis, which yields $ \vec{Q} = (0,0, \pi/c) $ and the primitive lattice vector interconnecting two sublattice points is $ (0,0,c) $, see Fig.~\ref{fig:structures}(a). The center of inversion lies in the middle between the two layers, e.g. at $ (0,0,c)/2 $ and interchanges the two sublattices ($ a$ and $ c$ being the lattice constants in-plane and out-of-plane, respectively). In this case, the subgroup leaving the sublattices invariant is $ {\cal  G}' = C_{4v} $ and $ \Gamma' =A_{2u} $. 
The second example is a sublattice structure within each layer with the primitive lattice vector $ (a,a,0) $ connecting the two sublattices, leading to $ \vec{Q} = (\pi/a, \pi/a,0) $, see Fig.~\ref{fig:structures}(b). The inversion center lies within the layer on a lattice point belonging to one of the two sublattices. The subgroup retaining the crystal structure is
$ C_{4h} $ and $ \Gamma' = A_{2g} $ of $ D_{4h} $. 

\begin{table}[bt]
    \centering
\begin{tabular}{c|cc|c}
     & intra-sublattice & inter-sublattice & IR\\
	 \hline
	 intra-band& $\tau^0$ & $\tau^3$ & $A_{1g}$\\
	 inter-band & $\tau^1$ & $\tau^{2}$ &$\Gamma'$
\end{tabular}
\caption{The different band dependencies possible for terms in the Hamiltonian of the systems under investigation here. While $\tau^0$ and $\tau^{3}$ always belong to the irreducible representation $A_{1g}$, the irreducible representation $\Gamma'$ of the other two Pauli matrices depends on the symmetry operations, that have to be combined with a sublattice interchange to map the crystal onto itself.}
    \label{tab:taus}
\end{table}

\subsection{Off-diagonal single-particle terms}

We now introduce the superconducting order parameter which on the mean-field level leads to off-diagonal terms to the single-particle Hamiltonian. These terms can be classified in a very analogous way as the diagonal terms. It is illustrative to discuss first the pair wavefunction
\begin{equation}
      \Psi_{\kk\alpha\alpha'}^{ss'} = \langle c_{\alpha \kk s}c_{\alpha'-\kk s'}\rangle,
    \label{eq:pairwave}
\end{equation}
which combines two electrons characterized by spin and band configuration. Note that the pair wavefunction describes zero-momentum pairs for $ \alpha = \alpha' $ while for 
$ \alpha \neq \alpha' $ the pairs possess momentum $ \Q $ as can be seen from the definition of the single-particle operators in Eq.~\eqref{eq:twoband}. 

\begin{table}[bt]
    \centering
\begin{tabular}{l|c|c}
$ \Gamma^+ $   & $ \psi_{0,1,3}(\kk) $ & $ \vec{d}_2(\kk) $ \\
 \hline
 $A_{1g} $ & 1 & $ \hat{x}k_yk_z - \hat{y}k_zk_x $  \\
 $A_{2g} $ & $ k_x k_y (k_x^2 -k_y^2) $ & $ \hat{x} k_x k_z + \hat{y} k_y k_z $    \\
 $B_{1g} $ & $ k_x^2 - k_y^2 $ & $ \hat{x}k_y k_z + \hat{y} k_x k_z $      \\
 $B_{2g} $ & $ k_x k_y $ & $ \hat{x} k_x k_z - \hat{y} k_y k_z $ \\
 $E_g $  &   $ \{ k_x k_z , k_y k_z \} $ &  $ \{ \hat{z} k_x k_z , \hat{z} k_y k_z     \} $\\[0.5 ex]
\hline \hline
 & & \\
 $ \Gamma^- $ & $ \psi_2(\kk) $ & $ \vec{d}_{0,1,3} (\kk) $ \\
 \hline
  $ A_{1u} $ & - &
 $ \hat{x} k_x + \hat{y} k_y + \epsilon \hat{z}k_z $ \\
$ A_{2u} $ & $ k_z $  & $ \hat{x}k_y - \hat{y} k_x $ \\
$ B_{1u} $ & $ k_x k_y k_z $  & $ \hat{x} k_x - \hat{y} k_y $ \\
$ B_{2u} $ & $ k_z (k_x^2 -k_y^2) $   & $ \hat{x}k_y + \hat{y} k_x $ \\
$ E_u     $  & $ \{ k_x , k_y \} $  & $ \{ \hat{z} k_x , \hat{z}k_y \} $ \\
 \end{tabular}
 \caption{Basis functions belonging to the different irreducible representations of D$_{4h}$ with SOC for the different gaps.}
    \label{tab:cooper-pair}
\end{table}

In order to formulate the off-diagonal terms in the Hamiltonian, we introduce now the (mean-field) gap function $\Delta_{\alpha\alpha'}^{ss'}(\kk)$ and write
\begin{equation}
  \HH_{\rm MF}' = \sum_{\kk}\Delta_{\alpha\alpha'}^{ss'}(\kk)c^{\dag}_{\alpha\kk s}c^{\dag}_{\alpha'-\kk s'} + {\rm h.c.}.
  \label{eq:mfterm}
\end{equation}
We use the standard notation of a scalar gap function $ \psi(\kk) $ for spin-singlet and the vector gap function $ \vec{d}(\kk) $ for spin-triplet pairing. The gap function has to satisfy the Pauli principle to change sign under exchange of the two electrons:
\begin{equation}
\Delta_{\alpha\alpha'}^{ss'}(\kk) = - \Delta_{\alpha'\alpha}^{s's}(-\kk) .
\label{pauli}
\end{equation}
For a single-band superconductor this requires that  $ \psi(-\kk) = \psi(\kk) $ and $ \vec{d}(-\kk) = - \vec{d}(\kk) $. 

We express the gap function as
\begin{equation}
\Delta^{ss'}_{\alpha \alpha'}(\kk,a) = [ \psi_a(\kk) \varsigma^0 + \vec{d}_a(\kk) \cdot \vec{\varsigma} ]_{ss'} \otimes \tau^a_{\alpha \alpha'} ,
\end{equation}
where we define $ \varsigma^0 = i \sigma^y $ and $ \vec{\varsigma} = i \vec{\sigma} \sigma^y $.
For intra-sublattice Cooper pairing originating from interactions between electrons on the same sublattice, 
intra-band corresponds to $a=0$ and inter-band pairing to $a=1$. 
Analogously, inter-sublattice pairing for intra-band pairs takes the index $ a=3 $ and for inter-band pairs $ a=2 $. Note that for $a = 0,1 $ and 3 the scalar (vector) gap function is an even (odd) function of $ \kk $, while it is opposite for $ a=2$, as required by Eq.~(\ref{pauli}). The case of $ a=2 $ is special in the sense that the band part of the pairing state
is antisymmetric under exchange allowing both momentum and spin part to be simultaneously symmetric or anti-symmetric. 

As in the case of the diagonal part we can classify the symmetry for the tensor product characterizing the pairing state (gap function). Thus, we consider again the 
irreducible representations $ R_s \otimes R_s' $ of the generating point group $ {\cal G} $. The representations $ R_s' $ correspond again to the ones of the $ \tau $ matrices 
as given in Tab.\ref{tab:taus}.  The representation $ R_s $ is based on the internal (spin and momentum) structure of the Cooper pair, given in Tab.\ref{tab:cooper-pair} for the
case $ {\cal G} = D_{4h} $, which we will use in the following. 

Together with the representations of the diagonal part it is possible to see which types of pairing states would couple together due to the staggered form of 
spin-orbit coupling. Such couplings occur due to combination of the diagonal and off-diagonal part in the linearized gap equations as a form of selection rules~\cite{dresselhaus:2008}. Thus, having the pairing state $ R_s \otimes R_s' $ also a state of symmetry $\tilde{R}_{s}\otimes\tilde{R}_{s}'$ would be coupled through a diagonal term of symmetry $ R \otimes R' $, if it appears in the decomposition of
\begin{equation}
(R \otimes R') \times  (R_s \otimes R_s') = (R \times R_s) \otimes ( R' \times R_s').
\label{eq:decomp}
\end{equation}
This allows to classify all possible interdependent pairing states within a given crystal lattice symmetry. 

In the following, we will analyze three different tetragonal crystal lattices with generating point group $ D_{4h} $ and elaborate on the way of analyzing the
influence of staggered types of spin-orbit coupling due to local inversion-symmetry breaking on superconductivity from a symmetry point of view.  


\section{Stack of Inversion Symmetry Lacking Layers}
\label{sec:layers}
Our first example is a tetragonal crystal lattice, whose staggered form originates from a sublattice structure of alternating layers. The basic unit, the layer, violates
inversion symmetry by the absence of reflection symmetry $ z \to -z $ ($z$: the four-fold rotation axis of the tetragonal crystal). This type of non-centrosymmetricity yields a Rashba-type spin-orbit coupling in each layer, 
\begin{equation}
  \HH^{\rm SOC} = \sum_{\kk}(\vec{\Lambda}_{\kk}\cdot\vec{\sigma}_{ss'})c^{\dag}_{\kk s}c^{\phantom{\dag}}_{\kk s'}
    \label{eq:soc1}
\end{equation}
with $ \vec{\Lambda}_{\kk} = \alpha (\hat{x} \sin k_y - \hat{y} \sin k_x) $. The sign of the Rashba coupling $ \alpha $ is opposite for the two sublattices, i.e.
alternates from layer to layer (see Fig.~\ref{fig:structures}(a)). The two bands resulting from this feature are related by $ \vec{Q} = (0,0,\pi) $ taking from now on all lattice constants to unity.

\label{sec:c1}
\subsection{Symmetry considerations}

The crystal lattice has the tetragonal $ D_{4h} $ point group with full inversion symmetry, taking the center at a symmetry point between the layers. The elements of $ D_{4h} $ are divided into those transforming within the layers and those interchanging the sublattice:

\begin{eqnarray}
G^{\rm intra} & = & \{ E, 2 C_4, C_2, 2 \sigma_v, 2 \sigma_d \} = C_{4v},   \\
G^{\rm inter} & = & \{ 2C_2' , 2C_2'', I, \sigma_h, 2 S_4\} ,
\end{eqnarray}
using the standard notation of Ref.~\onlinecite{Landau}. From this we conclude that $ R' = A_{2u} $, which is the one-dimensional irreducible representation with $+1$ for all elements of $ G^{\rm intra} $ and $-1$ for all elements of  $ G^{\rm inter} $. Considering now the diagonal single-particle part of the Hamiltonian, we find for the spin-independent hopping terms the standard representations $ A_{1g} \otimes A_{1g} $.
On the other hand, the staggered spin-orbit part consists only of the intra-sublattice (in-plane) Rashba-like 
coupling for which $ \vec{g}_{\kk} $ transforms according to $ A_{2u} $ and corresponds to $ \vec{\Lambda}_{\vec{k}} = \vec{\zeta}_{\kk}^{\rm intra} $ of Eq.~(\ref{f1-g2}). This leads to the representation $ A_{2u} \otimes A_{2u} $. 

For this system, spin-orbit coupling mixes pairing states in the following way,
\begin{equation}
(A_{2u} \otimes A_{2u} ) \times (R_s \otimes R_{s'} ) = (A_{2u} \times R_s) \otimes (A_{2u} \times R_s'),
\end{equation}
i.e. states of opposite parity can be mixed, as is generally the case in non-centrosymmetric systems. Note that this also implies that
intra-band pairs mix with inter-band pairs. 

Looking first at intra-sublattice (intra-layer) pairing states, we consider the example of the (even-parity) s-wave spin-singlet state, which has for
intra-band pairing the representation $ A_{1g} \otimes A_{1g} $ while it belongs to $ A_{1g} \otimes A_{2u} $ for inter-band pairing. 
The mixing occurs as follows,
\begin{equation} \begin{array}{l}
A_{1g} \otimes A_{1g} \leftrightarrow A_{2u} \otimes A_{2u} , \\
A_{1g} \otimes A_{2u} \leftrightarrow A_{2u} \otimes A_{1g} ,
\end{array}
\label{example-1}
\end{equation}
whereby the admixed states have always opposite parity ($ A_{2u} $). 
Using Tab.\ref{tab:cooper-pair} we write the two types of states with intra-sublattice pairing as
\begin{equation} \begin{array}{ll}
\hat{\Delta} (\kk) & = \psi_0 \varsigma^0\otimes \tau^0 + d_1 (k_y \varsigma^x - k_x \varsigma^y)\otimes \tau^1, \\
\hat{\Delta} (\kk) & = \psi_1 \varsigma^0\otimes \tau^1 + d_0 (k_y \varsigma^x - k_x \varsigma^y)\otimes \tau^0,
\end{array}
\label{pair-intra-sub}
\end{equation} 
which mix the spin-singlet and triplet configurations. Note that the same scheme also applies for other pairing states, e.g., a d-wave state beloning to $B_{1g}\otimes A_{1g}$ couples to a spin-triplet pairing state belonging to $B_{2u}\otimes A_{2u}$. 

Next we consider inter-sublattice (inter-layer) pairs, starting with s-wave intra-band states, corresponding again to $ A_{1g} \otimes A_{1g} $
with the admixed $ A_{2u} \otimes A_{2u} $. On the other hand, the inter-band (even-parity) ``s-wave'' state ($ A_{1g} \otimes A_{2u} $) has a
spin-triplet configuration and couples to the intra-band odd-parity spin-triplet state $ A_{2u} \otimes A_{1g} $ as the $ \tau^2 $ matrix 
is involved (Tab.\ref{tab:taus}).

For the two possible inter-sublattice pairing states we find the gap functions
\begin{equation}
\hat{\Delta} (\kk) = \psi_3 \varsigma^0\otimes \tau^3 + \psi_2 k_z  \varsigma^0\otimes \tau^2 , \\
\end{equation}
and
\begin{multline}
\hat{\Delta} (\kk) = d_2 (k_yk_z \varsigma^x - k_xk_z  \varsigma^y)\otimes \tau^2 \\
 + d_3 (k_y \varsigma^x - k_x \varsigma^y)\otimes \tau^3,
\label{pair-inter-sub}
\end{multline}
which remain in either the spin-singlet or spin-triplet channel. Due to the sublattice structure, however, always inter- and intra-band states are mixed.

\subsection{Microscopic consideration}

To illustrate the symmetry-based aspects from a microscopic point of view we introduce here a model based on a tight-binding band structure, whereby each layer is considered as a simple square lattice. We use the two-band formulation and write the single-particle part of the Hamiltonian as
\begin{equation}
  \HH = \sum_{\kk} (\Xi_{\kk\alpha\alpha'}\sigma^0_{ss'} + \Gamma_{\kk\alpha\alpha'}^{ss'}) c_{\alpha\kk s}^{\dag}c_{\alpha'\kk s'}^{\phantom{\dag}}
    \label{eq:genham}
\end{equation}
with electron operators as defined in Eq.~\eqref{eq:twoband} with $\Q = (0, 0, \pi)$. Intra-layer hopping is taken into account between nearest and next-nearest neighbors and inter-layer only between nearest neighbors, which leads to 
\begin{equation} \begin{array}{ll}
\varepsilon_{\kk}^{\rm intra} & = -2t_{xy} (\cos k_x + \cos k_y) - 4t_{xy}'\cos k_x \cos k_y , \\
\varepsilon_{\kk}^{\rm inter} & = -2t_z \cos k_z .
\end{array}
\end{equation}
These contribute to the spin-independent part. The spin-dependent part originates from the staggered Rashba-type spin-orbit coupling, which we take only in the nearest-neighbor form as in Eq.~(\ref{eq:soc1}) yielding
\begin{equation}
  \vec{f}_{\kk}^{\,1} = \alpha(\hat{x}\sin k_y - \hat{y}\sin k_x)
  \label{eq:fk1}
\end{equation}
following Eqs.~\eqref{eq:Gammaintra} and \eqref{eq:Gammainter}.  

It is convenient for the following to use the formulation by means of Green's functions, which for the non-interacting case can straightforwardly be calculated 
by inverting the $(4\times4)$ matrix $(i\omega_n \sigma_{ss'}^0\otimes\tau_{\alpha \alpha'}^0 -\Xi_{\kk\alpha\alpha'}\sigma^{0}_{ss'} - \Gamma_{\kk\alpha\alpha'}^{ss'})$,
\begin{multline}
    \hat{G}_0(\kk, \omega_n) = G_{0+}(\kk, \omega_n)\sigma^0\otimes\tau^0\\
    +G_{0-}(\kk, \omega_n)(\hat{f}_\kk\cdot\vec{\sigma}\otimes\tau^1 + \hat{\varepsilon}_{\kk}\sigma^0\otimes\tau^3),
    \label{eq:c1-greens0}
\end{multline}
where
\begin{equation}
    G_{0\pm}(\kk, \omega_n) = \frac{1}{2}\Big(\frac{1}{i\omega_n - \xi_{+, \kk}}\pm \frac{1}{i\omega_n - \xi_{-, \kk}}\Big)
    \label{eq:c1-greens0pm},
\end{equation}
\begin{equation}
  \hat{f}_{\kk} = \vec{f}_{\kk}^{\,1}/\sqrt{|\vec{f}_\kk^{\,1}|^2 + (\varepsilon_{\kk}^{\rm inter})^2}
\end{equation}
and 
\begin{equation}
  \hat{\varepsilon}_{\kk} = \varepsilon_{\kk}^{\rm inter}/\sqrt{|\vec{f}_\kk^{\,1}|^2 + (\varepsilon_{\kk}^{\rm inter})^2}.
\end{equation}
In Eq.~\eqref{eq:c1-greens0pm}, the two (spin-independent) band energies are given by
\begin{equation}
  \xi_{\pm, \kk s} = \xi_{\pm, \kk} = \varepsilon_{\kk}^{\rm intra} - \mu \pm \sqrt{|\vec{f}_{\kk}^{\,1}|^2+ (\varepsilon_{\kk}^{\rm inter})^2}. 
    \label{eq:sym-c1-energies0}
\end{equation}

We now turn to the problem of superconductivity by introducing a pairing interaction of the general form, 
\begin{equation}
    \HH'=\frac{1}{N}\sum_{\kk, \kk'}V_{\alpha\beta, \mu\nu}^{ss', s_3s_4}(\kk, \kk')c_{\alpha \kk s}^{\dag}c_{\beta-\kk s'}^{\dag}c_{\mu-\kk's_3}^{\phantom{\dag}}c_{\nu \kk's_4}^{\phantom{\dag}} .
    \label{eq:int}
\end{equation}
We parametrize the matrix element in the notation used for the single-particle terms,
\begin{multline}
  V_{\alpha\beta, \mu\nu}^{ss', s_3s_4}(\kk, \kk')=\sum_{m,n}\sum_{a}v_{mn}^{(a)}[\psi_{mn}^{(a)}(\kk)\varsigma^{m}_{ss'}\tau^{n}_{\alpha\beta}]\\
  \times[\psi_{mn}^{(a)}(\kk')\varsigma^{m}_{s_{3}s_{4}}\tau^{n}_{\mu\nu}]^{\dag},
	\label{eq:c1-pairing}
\end{multline}
where $\psi_{mn}^{(a)}(\kk)$ have the symmetry of the gap functions tabulated in Tab.\ref{tab:cooper-pair}. For a more detailed analysis of the structure of such an interaction see appendix~\ref{app:interaction}. This pairing interaction incorporates both, coupling of the intra- and inter-sublattice type. For simplicity, we restrict ourselves to interactions including only nearest-neighbor coupling in the real lattice. This limits the classification of pairing states as can be seen in Tab.\ref{tab:c1-gapfunctions} compared to the more general
Tab.\ref{tab:cooper-pair}. 

With the Hamiltonian and the non-interacting Green's function introduced above it is possible to analyze the superconducting instabilities in detail by resorting to the standard framework of the Gor'kov equations\cite{Mineev}. The linearized gap equation reads
\begin{multline}
	\Delta_{\alpha\beta}^{ss'}(\kk)= -T  \sum_{\mu,\nu}\sum_{\omega_n}\sum_{\kk'} \sum_{s_3,s_4}  V_{\alpha\beta,\mu\nu}^{ss's_3s_4}(\kk, \kk')\\
	\times [\hat{G}_0(\kk', \omega_n) \hat{\Delta}(\kk') \hat{G}_0^T(-\kk', -\omega_n)]_{\nu\mu}^{s_4s_3}, 
    \label{eq:lingap}
\end{multline}
where all the Green's functions as well as the order parameter are $4\times4$ matrices. This gap equation is analyzed in the following for the two cases of a leading instability in the intra-sublattice and the inter-sublattice pairing channel, respectively.

\subsubsection{Intra-layer interaction}

We use the nearest-neighbor interactions derived in appendix~\ref{app:interaction} for the intra-sublattice case, which, following Eq.~(\ref{pair-intra-sub}), lead to gap functions of the form
\begin{equation}
    \hat{\Delta}(\kk) = \left\{ \begin{array}{l} \psi_{0}(\kk) \varsigma^{0}\otimes \tau^{0} +  \vec{d}_{1}(\kk)\cdot \vec{\varsigma} \otimes \tau^{1} \\
    \psi_{1}(\kk) \varsigma^{0} \otimes \tau^{1} + \vec{d}_{0}(\kk)\cdot \vec{\varsigma} \otimes \tau^{0}
    \end{array} \right.
    \label{eq:c1-intragap}
\end{equation}
for which we insert from Tab.\ref{tab:c1-gapfunctions}
\begin{eqnarray} 
\psi_n (\kk) &=&  \psi_n (\cos k_x + \cos k_y),\\ 
\vec{d}_n (\kk) &=&  d_n (\hat{x} \sin k_y - \hat{y} \sin k_x) 
\end{eqnarray}
with $ n = 0, 1 $. It is easy to see that the gap functions \eqref{eq:c1-intragap} couple within the linearized gap equation (\ref{eq:lingap}) indeed in the way anticipated above,
using the intra-layer interaction given in the appendix in Eqs.~\eqref{eq:plus} and \eqref{int-minus}: 
\begin{multline}
  \psi_{0}(\kk) = -T\sum_{n,\kk'}v^{+}_{\kk \kk'}\Big\{[G_{0+}\tilde{G}_{0+} + G_{0-}\tilde{G}_{0-}]\psi_{0}(\kk')\\
    + [G_{0+}\tilde{G}_{0-}+G_{0-}\tilde{G}_{0+}]\hat{f}_{\kk'}\cdot\vec{d}_{1}(\kk')\Big\},\label{eq:c1-3d-singlet1}
\end{multline}
\begin{multline}
  \vec{d}_{1}(\kk) = -T\sum_{n,\kk'}v^{-}_{\kk \kk'}\Big\{[G_{0+}\tilde{G}_{0+} + G_{0-}\tilde{G}_{0-} ]\vec{d}_{1}(\kk')\\
     +2G_{0-}\tilde{G}_{0-} \{ \hat{f}_{\kk'} [\hat{f}_{\kk'}\cdot\vec{d}_{1}(\kk')] -\vec{d}_{1}(\kk')\} \\
     + [G_{0+}\tilde{G}_{0-}+G_{0-}\tilde{G}_{0+}]\hat{f}_{\kk'}\psi_{0}(\kk')\Big\}
    \label{eq:c1-3d-singlet}
\end{multline}
and, analogously,
\begin{multline}
  \vec{d}_{0}(\kk) = -T\sum_{n,\kk'}v^{-}_{\kk \kk'}\Big\{[G_{0+}\tilde{G}_{0+} + G_{0-}\tilde{G}_{0-}]\vec{d}_{0}(\kk')\\
    + 2G_{0-}\tilde{G}_{0-}\{\hat{f}_{\kk'} [\hat{f}_{\kk'}\cdot \vec{d}_{0}(\kk')] - \vec{d}_{0}(\kk')\}\\
    +[G_{0+}\tilde{G}_{0-}+G_{0-}\tilde{G}_{0+}]\hat{f}_{\kk'}\psi_{1}(\kk')\Big\},
\end{multline}
\begin{multline}
  \psi_{1}(\kk) = -T\sum_{n,\kk'}v^{+}_{\kk \kk'}\Big\{[G_{0+}\tilde{G}_{0+} +G_{0-}\tilde{G}_{0-}]\psi_{1}(\kk')\\
    -2 (\hat{\varepsilon}_{\kk'})^{2}G_{0-}\tilde{G}_{0-}\psi_{1}(\kk')\\
     + [G_{0+}\tilde{G}_{0-}+G_{0-}\tilde{G}_{0+}]\hat{f}_{\kk'}\cdot\vec{d}_{0}(\kk')\Big\}.
    \label{eq:c1-3d-triplet}
\end{multline}
Here, we have introduced the short notation $G_{0\pm} = G_{0\pm}(\kk, \omega_{n})$ and $\tilde{G}_{0\pm} = G_{0\pm}(-\kk, -\omega_n)$. 
It is also obvious now that it is the spin-orbit coupling term, represented here through $ \hat{f}_{\kk} $, which yields the
coupling between even- and odd-parity pairing states. 

\begin{table}[bt]
    \centering
\begin{tabular}{c|c|c}
  & \;\;\; intra-sublattice \;\;\;& \;\;\;inter-sublattice\;\;\;\\
	 \hline
	 \hline
	 & $\psi_{0, 1}(\kk)$ & $\psi_{3}(\kk)$  \\[0.5ex]
	 \hline
	 \;\;$A_{1g}$ \;\;& $1$, $\cos k_x + \cos k_y$ & $\cos k_z$\\
	 $B_{1g}$ & $\cos k_x - \cos k_y$ & \\[0.5ex]
	 \hline
	 \hline
	 & $\vec{d}_{0,1}(\kk)$ &  $\psi_{2}(\kk) , \vec{d}_3(\kk) $ \\[0.5ex]
	 \hline
	 $A_{1u}$ & $\hat{x}\sin k_x + \hat{y}\sin k_y$ &$ \hat{z} \sin k_z$ \\
	 $A_{2u}$ & $\hat{x}\sin k_y - \hat{y}\sin k_x$ & $\sin k_z$\\
	 $B_{1u}$ & $\hat{x}\sin k_x - \hat{y}\sin k_y$ & \\
	 $B_{2u}$ & $\hat{x}\sin k_y + \hat{y}\sin k_x$ & \\
	  $E_u    $ & $ \{ \hat{z} \sin k_x, \hat{z} \sin k_y \} $ & 
\end{tabular}
\caption{List of different basis functions for a crystal structure with an alternating stack of mirror-symmetry lacking layers that are supported by nearest-neighbor intra-sublattice and inter-sublattice interaction, respectively. Note that with the restriction that only nearest-neighbor pairing is considered the ``inter-sublattice'' case does not include any even-parity spin-triplet $ \vec{d}_2 (\kk) $-states.}
    \label{tab:c1-gapfunctions}
\end{table}

\subsubsection{Inter-layer interaction}

Turning to the inter-sublattice (inter-layer) pairing the situation becomes more intricate due to pairing in the anti-symmetric band channel ($ \tau^2$). We write the
gap function as
\begin{equation}
 \hat{\Delta}(\kk) = \left\{ \begin{array}{l} \psi_{2}(\kk) \varsigma^{0}\otimes \tau^{2} +  \vec{d}_{3}(\kk)\cdot \vec{\varsigma} \otimes \tau^{3} \\
    \psi_{3}(\kk) \varsigma^{3} \otimes \tau^{3} + \vec{d}_{2}(\kk)\cdot \vec{\varsigma} \otimes \tau^{2}
    \end{array} \right.,
    \label{eq:c1-intergap}
\end{equation}
where for nearest-neighbor pairing only we may use
\begin{eqnarray}
\psi_2 (\kk) &=&  \psi_2 \sin k_z,\\
\psi_3 (\kk) &=&  \psi_3 \cos k_z,\\
\vec{d}_3 (\kk) &=&  d_3 \hat{z} \sin k_z,
\end{eqnarray}
while $ \vec{d}_2 (\kk) = 0 $, following Tab.\ref{tab:c1-gapfunctions}. To write the linearized gap equation we use the inter-layer pairing interactions (\ref{eq:internnp}) and (\ref{eq:internnm}) to find 
\begin{multline}
  \psi_{3}(\kk) = -T\sum_{n,\kk'}v^{+}_{\kk \kk'} \left\{  [ G_{0+}\tilde{G}_{0+} +G_{0-}\tilde{G}_{0-}] \psi_{3}(\kk') \right. \\
 \left.    - 2  G_{0-}\tilde{G}_{0-} \hat{f}^{2}_{\kk'} \psi_{3}(\kk') \right\}
    \label{eq:c1-even-singlet}
\end{multline}
and, in the same way,
\begin{multline}
  \vec{d}_{3}(\kk) = -T\sum_{n,\kk'}v^{-}_{\kk \kk'}\Big\{[G_{0+}\tilde{G}_{0+} + G_{0-}\tilde{G}_{0-}]\vec{d}_{3}(\kk')\\
    -2G_{0-}\tilde{G}_{0-}[\hat{f}_{\kk'}\cdot\vec{d}_{3}(\kk')]\hat{f}_{\kk'}\Big\}.
    \label{eq:c1-odd-triplet}
\end{multline}
Note that for the gap equation there is no mixing within this approximation. 

\subsection{Discussion}
\begin{figure}[tb]
    \begin{center}
	\includegraphics{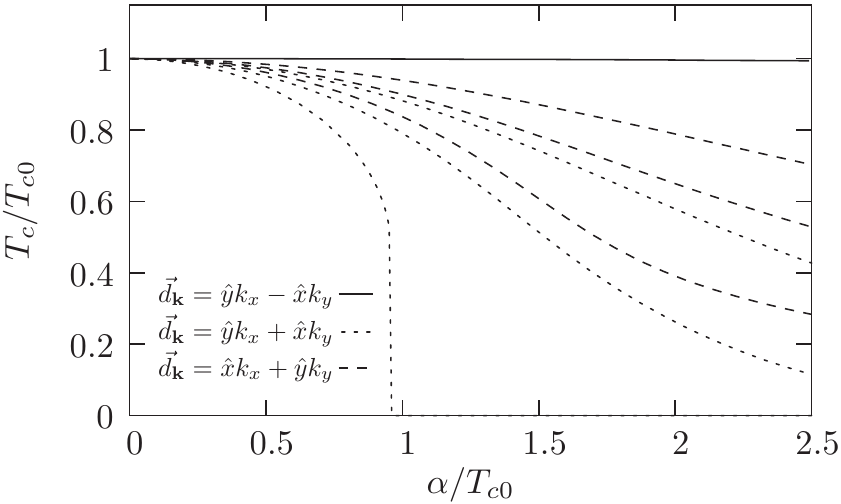}
    \end{center}
    \caption[Suppression of $T_c$ due to antisymmetric SOC]{Suppression of the transition temperature of the intra-band gaps due to the antisymmetric SOC. With increasing inter-layer coupling, the suppression is weakened, $t_{z}=0, 0.1t, 0.2t$ from bottom to top.}
    \label{fig:sym-ta}
\end{figure}

We first consider the intra-layer pairing channels looking at Eqs.~(\ref{eq:c1-3d-singlet1})-(\ref{eq:c1-3d-triplet}). Restricting our analysis to the terms diagonal in each gap function only, we find that even-parity intra-band pairing is essentially unaffected by spin-orbit coupling. More interesting are the intra-band odd-parity gaps which suffer suppression unless the $d$ vector is parallel to $ \hat{f}_{\kk} $. This property is known from non-centrosymmetric superconductors \cite{frigeri:2004}
and should be fully transferable to the case of completely decoupled layers. Our analysis shows that including inter-band pairing combined with inter-layer hopping ($t_z \neq 0 $) reduces the pair breaking effect of spin-orbit coupling, as can be seen in Fig.~\ref{fig:sym-ta}. There, we plot $ T_c$ versus the spin-orbit coupling strength $ \alpha $ for three different values of the inter-layer hopping values. Naturally, the case of $ \vec{d}_0(\kk) \parallel \hat{f}_{\kk} $ is unchanged (solid line in Fig.~\ref{fig:sym-ta}). 

\begin{figure}[tb]
    \begin{center}
	\includegraphics{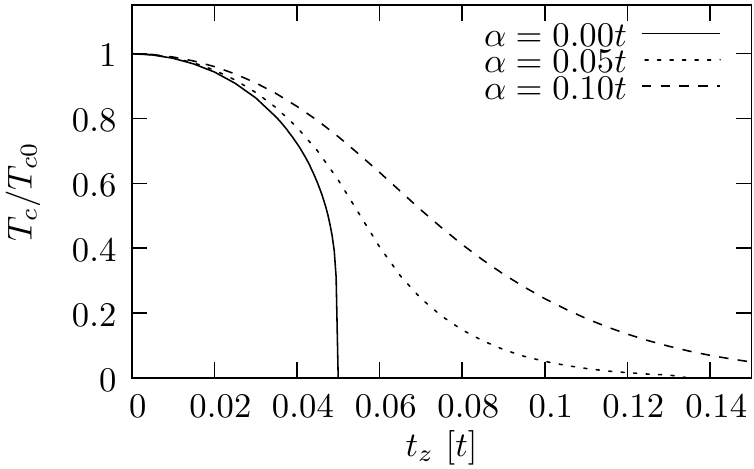}
    \end{center}
    \caption[Suppression of $T_c$ due to inter-layer hopping]{Change in the transition temperature of the inter-band gap with $\vec{d}_{1}(\kk) = \hat{x}\sin k_{y}-\hat{y}\sin k_{x}$ as a function of the inter-layer hopping $t_{z}$ for different spin-orbit coupling strengths.}
    \label{fig:sym-ttz}
\end{figure}

Considering the inter-band gaps [$\psi_1(\kk)$ and $\vec{d}_{1}(\kk)$], which correspond to finite-momentum pairing, it is easy to see that they have the same gap equations as the intra-band gaps, if we decouple the layers ($ t_z = 0 $ with $ \hat{\varepsilon}_{\kk} = 0 $) and set the spin-orbit coupling to zero ($ \alpha = 0 $ with $ \hat{f}_{\kk} = 0 $). The reason is that the finite momentum  $ \vec{Q} = (0,0,\pi) $ yields an alternating phase of $ 0 $ and $ \pi $ from layer to layer, which is irrelevant for decoupled layers. Therefore finite-momentum pairing is
suppressed by inter-layer hopping, which introduces the disadvantage of an inter-layer phase shift to the energy balance, 
as we can see in the plot of $ T_c $ in Fig.~\ref{fig:sym-ttz}. On the other hand, adding spin-orbit coupling helps the intra-band pairing to slightly recover $ T_c $. 

The trends discussed so far show that a strong inter-layer coupling moves the system further away from the parity-mixing (spin singlet-triplet mixing) as compared to a real non-centrosymmetric superconductor. Inter-layer coupling ``recovers the inversion symmetry'' gradually. 

A further interesting aspect occurs, if the inter-layer pairing is the leading instability. While the system has the full inversion symmetry in this case, the spin-orbit coupling acts as pair-breaking for the spin-singlet channel (see Eq.~(\ref{eq:c1-even-singlet})). On the other hand, spin-orbit coupling does not affect the spin-triplet gap $ \vec{d}_3(\kk) $,
since $ \vec{d}_3(\kk) \perp \hat{f}_{\kk} $ for all $ \kk $. Thus, in this case the spin-orbit coupling can be important in 
influencing the pairing symmetry in favor of a spin-triplet state. Note, however, that the structure of the pairing interaction remains the major deciding element for the pairing symmetry. 

\section{Inversion Symmetry Lacking Bonds}
\label{sec:bonds}
The second example studied here is motivated by the layered perovskite crystal structure with tetragonal symmetry, known for some transition metal oxides. The subunits are oxygen octahedra, where six oxygen ions enclose a transition metal ion. In some systems, these octahedra rotate around the crystalline $z$ axis leading to a staggered pattern of rotation, i.e. neighboring octahedra in the $xy$ plane rotate in opposite direction [see Fig.~\ref{fig:structures}(b)]. Such features are known in the bilayer Sr$_3$Ru$_2$O$_7$ and in Sr$_2$RhO$_4$ or Sr$_2$IrO$_4$, to mention a few examples.~\cite{shaked:2000,subramanian:1994} 
This lattice distortion shifts the in-plane bond oxygens to off-center positions, and thus, leads to a breaking of inversion symmetry on each of these bonds. Again, we arrive at a form of staggered spin-orbit coupling fitting into the scheme developed above. Since inversion symmetry within the sublattice is retained in this structure, there is no even-odd mixing at all for this type of crystal structure. However, the spin-orbit coupling has an influence on the direction of the $d$-vector for a spin-triplet
pairing state. Before this is studied in detail with the help of the linearized gap equation, we again start with a symmetry analysis. 

\subsection{Analysis of symmetry}
Without the rotation of the O$_{6}$ octahedra, the point group of a single layer is $D_{4h}$ as in the above example. With the rotation, the unit cell is doubled in plane as depicted by the light and dark lattice sites in Fig.~\ref{fig:structures}(b). The corresponding $ Q$-vector is $ \vec{Q} = (\pi,\pi,0) $.

We separate again the symmetry operations within $D_{4h} $ which turn each sublattice into itself ($ G^{\rm intra} $)  while the remaining operations ($G^{\rm inter} $) exchange the sublattices,
\begin{equation} \begin{array}{l}
G^{\rm intra} = \{ E, 2 C_4, C_2,I, 2 S_4 , \sigma_h \} = C_{4h} , \\ \\
G^{\rm inter} = \{2 C_2', 2C_2'', 2 \sigma_v, 2 \sigma_d \} .
\end{array} \end{equation}
The representation of $ D_{4h} $ changing sign for all elements of $ G^{\rm inter} $ is $ \Gamma' = A_{2g} $. 

Analogous to the previous example, the terms in the Hamiltonian can be characterized with respect to their behavior under sublattice interchange. According to the above symmetry analysis, the terms that change sign belong to the irreducible representation $\Gamma'= A_{2g}$ for this structure and a symmetry-reducing term in the Hamiltonian has to be of $A_{2g}\otimes A_{2g}$ symmetry. The staggered spin-orbit coupling derived in Ref.~\onlinecite{fischer:2010} is of this symmetry with
\begin{equation}
  \vec{g}_{\kk}^{\,2} =2\alpha(\cos k_x + \cos k_y)\hat{z} = \vec{\zeta}_{\kk}^{\rm inter}
    \label{eq:c3-soc}
\end{equation}
and has even parity.

It is again possible to find the symmetry allowed couplings between different gap functions by factorizing the gap function in terms of spin and orbital degrees of freedom to find the general relation for mixing pairing channels,
\begin{equation}
\Gamma \otimes A_{1g} \leftrightarrow (\Gamma \times A_{2g}) \otimes A_{2g},
\end{equation}
which, for instance, leads to
\begin{eqnarray}
    A_{1g}\otimes A_{1g} &\leftrightarrow& A_{2g}\otimes A_{2g},\label{eq:sym-c1-couplings1}\\
    B_{1g,u}\otimes A_{1g} &\leftrightarrow& B_{2g,u}\otimes A_{2g},\label{eq:sym-c1-couplings2}\\
    E_{g,u}\otimes A_{1g} &\leftrightarrow& E_{g,u}\otimes A_{2g}.
    \label{eq:c3-couplings}
\end{eqnarray}
As mentioned above, spin-orbit coupling here does not mix states of different parity. 
\subsection{Analysis of instability}
A better understanding of the consequence of symmetry properties and of the influence of the spin-orbit coupling on the different superconducting states can be obtained by analyzing the linearized self-consistency equation for the gap~\eqref{eq:lingap}. The non-interacting Hamiltonian is the same as in Section~\ref{sec:layers} with the only difference that the spin-dependent term here uses $\vec{g}_{\kk}^{\,2}$ as given in Eq.~\eqref{eq:c3-soc}. Note that we restrict ourselves to the single-band case 
and ignore the aspect of degenerate $d$ orbitals of transition metal ions in the examples mentioned above. 

The non-interacting Green's function is given by
\begin{multline}
    G_0(\kk, \omega_n) = G_{0+}(\kk, \omega_n)\sigma^0\otimes\tau^0\\
    - G_{0-}(\kk, \omega_n)(\hat{g}_\kk\sigma^z\otimes\tau^2 - \hat{\varepsilon}_{\kk}\sigma^0\otimes\tau^3),
    \label{eq:c3-greens0}
\end{multline}
with
\begin{equation}
    G_{0\pm}(\kk, \omega_n) = \frac{1}{2}(\frac{1}{i\omega_n - \xi_{+, \kk}}\pm \frac{1}{i\omega_n - \xi_{-, \kk}})\label{eq:c3-greens0pm},
\end{equation}
\begin{equation}
  \hat{g}_{\kk} = (\vec{g}_{\kk}^{\, 2})_z/\sqrt{|\vec{g}_\kk^{\, 2}|^2 + (\varepsilon_{\kk}^{\rm inter})^2},
\end{equation}
\begin{equation}
  \hat{\varepsilon}_{\kk} = \varepsilon_{\kk}^{\rm inter}/\sqrt{|\vec{g}_\kk^{\, 2}|^2 + (\varepsilon_{\kk}^{\rm inter})^2},
\end{equation}
and
\begin{equation}
  \xi_{\pm, \kk s} = \xi_{\pm, \kk} = \varepsilon_{\kk}^{\rm intra} - \mu \pm \sqrt{|\vec{g}_{\kk}^{\, 2}|^2+ (\varepsilon_{\kk}^{\rm inter})^2}. 
    \label{eq:c3-energies0}
\end{equation}
Note that we again distinguish hoppings connecting different sublattices (inter-sublattice, $\varepsilon_{\kk}^{\rm inter}$) and the same sublattice (intra-sublattice $\varepsilon_{\kk}^{\rm intra}$) with the former including nearest-neighbor and the latter including next-nearest-neighbor hopping. 
In the following, the two cases of an intra-sublattice and inter-sublattice pairing interaction are again discussed separately.  
\subsubsection{Intra-sublattice pairing}
For a leading interaction of intra-sublattice type, the gap is analogous to the form given in Eq.~\eqref{eq:c1-intragap}. 
As mentioned, the intra- and inter-band gap functions have the same parity and spin configuration. The case of
even-parity pairing can be illustrated with the example of the intra-band pairing state
\begin{equation}
\psi_0(\kk) = \sin k_x \sin k_y
\end{equation}
belonging to $ B_{2g} $ of $D_{4h} $. The corresponding admixed state according to Eq.~\eqref{eq:sym-c1-couplings2} is the $ B_{1g} $-state 
\begin{equation}
\psi_1 (\kk) = \cos 2k_x - \cos 2 k_y
\end{equation}
Note that $ \psi_0(\kk) $ here is based on next-nearest-neighbor pairing, while $ \psi_1(\kk) $ originates from
an interaction on sites separated by lattice vectors $ (2a,0) $.  

For the odd-parity channel it turns out that 
the $x$ and $y$ components of the $d$ vector for intra- and inter-band pairing states mix in the form
\begin{equation}
\left( \begin{array}{cc} d^x_0 (\kk) \\ d^y_0 (\kk) \end{array} \right) \leftrightarrow \left( \begin{array}{cc} - d^y_1 (\kk) \\ d^x_1 (\kk) \end{array} \right) ,
\end{equation}
which is a result of the decomposition of $ \Gamma \otimes A_{2g} $. This 
leads to combinations of gap functions like
\begin{multline}
A_{1u}:  \vec{d}_0 (\kk) = \hat{x} \sin (k_x + k_y) + \hat{y} \sin(k_x-k_y) \leftrightarrow\\ 
A_{2u}: \vec{d}_1(\kk) = \hat{x} \sin(k_x -k_y) - \hat{y} \sin (k_x + k_y) ,
\end{multline}
representing one example of a pairing state classified within the representation of $ D_{4h} $ and arising from next-nearest-neighbor interaction. The $z$ component of the $d$ vector is conserved in the mixing of inter- and intra-band pairing. This is fully compatible with the classification of spin-triplet pairing states in a tetragonal crystal lattice. 
The state belonging to $E_u $ then yields for next-nearest-neighbor pairing,
\begin{equation}
\vec{d}_0 (\kk) = \{ \hat{z} \sin (k_x + k_y) , \hat{z} \sin (k_x - k_y) \} 
\end{equation}
mixing with 
\begin{equation}
\vec{d}_1 (\kk) = \{ -\hat{z} \sin (k_x - k_y) , \hat{z} \sin (k_x + k_y) \},
\end{equation}
which also lies in the representation $E_u$.

\subsubsection{Inter-sublattice pairing}
\begin{table}[bt]
    \centering
    \begin{tabular}{l|c|c}
      $ \Gamma^+ $   & $ \psi_{3}(\kk) $ & $ \vec{d}_2(\kk) $ \\
      \hline
      $A_{1g} $ & $\cos k_x + \cos k_y$  &  -   \\
      $A_{2g} $ &  -   & $ \hat{z}(\cos k_x + \cos k_y) $    \\
      $B_{1g} $ & $ \cos k_x - \cos k_y $ & -      \\
      $B_{2g} $ & - & $ \hat{z} (\cos k_x - \cos k_y) $ \\
      $E_g $  &  - &  $ \{ \hat{x} (\cos k_x+ \cos k_y) , \hat{y} (\cos k_x + \cos k_y) \} $\\[0.5 ex]
      \hline\hline
      & & \\
      $ \Gamma^- $ & $ \psi_2(\kk) $ & $ \vec{d}_{3} (\kk) $ \\
      \hline
      $ A_{1u} $ & - & $ \hat{x} \sin k_x + \hat{y} \sin k_y $ \\
      $ A_{2u} $ & - & $ \hat{x} \sin k_y - \hat{y} \sin k_x $ \\
      $ B_{1u} $ & - & $ \hat{x} \sin k_x - \hat{y} \sin k_y $ \\
      $ B_{2u} $ & - & $ \hat{x} \sin k_y + \hat{y} \sin k_x $ \\
      $ E_u    $ & $ \{ \sin k_x , \sin k_y \} $  & $ \{ \hat{z} \sin k_x , \hat{z}\sin k_y \} $ \\
    \end{tabular}
    \caption{Basis functions belonging to the different irreducible representations of D$_{4h}$ supported by in-plane nearest-neighbor interactions. }
    \label{tab:c2-basis}
\end{table}
  
For inter-sublattice pairing interactions, we again consider a gap of the form given in Eq.~\eqref{eq:c1-intergap}. If we only consider pairing within the $xy$ plane, the spin-singlet pairing states $ \psi_3 (\kk) $ only appear in the one-dimensional representations of $D_{4h} $, while the states $\psi_2(\kk) $ are in the two-dimensional representation $E_u$ (the respective others require that the gap function changes sign under the operation $ z \to - z$). Therefore, following Table~\ref{tab:c2-basis} we find that corresponding spin-triplet components $ \vec{d} (\kk) \perp \hat{z} $ remain independent. Only the $E_{g,u} $ spin-triplet state ($ \vec{d}(\kk) \parallel \hat{z} $) mixes with the spin-singlet states. 

We consider first the case $ \vec{d} (\kk) \perp \hat{z} $ yielding the following linearized gap equation, 
\begin{equation}
  d_{3}^{x,y}(\kk) = -T\sum_{n, \kk'}4v^{-}_{\kk \kk'}(G_{0+}\tilde{G}_{0+} + G_{0-}\tilde{G}_{0-})d_{3}^{x,y}(\kk')
    \label{eq:sym-c1-eo-xy2}
\end{equation}
and
\begin{equation}
  d_{2}^{x,y}(\kk) = -T\sum_{n, \kk'}4v^{+}_{\kk \kk'}(G_{0+}\tilde{G}_{0+} - G_{0-}\tilde{G}_{0-})d_{2}^{x,y}(\kk'),
    \label{eq:sym-c1-eo-xy1}
\end{equation}
where we used again the short-hand notation $G_{0\pm}=G_{0\pm}(\kk', \omega_n)$ and $\tilde{G}_{0\pm}=G_{0\pm}(-\kk', -\omega_n)$ and $v^{\pm}_{\kk \kk'}$ are defined in App.~\ref{app:interaction}. 

On the other hand, the $z$ component mixes with a scalar gap function, 
\begin{equation}
  \left(\begin{array}{c} d_{3}^{z}(\kk) \\ \psi_{2}(\kk)\end{array}\right) = -T\sum_{n, \kk'}4v^{-}_{\kk \kk'}\left[M(\kk')\right]\left(\begin{array}{c} d_{3}^{z}(\kk')\\ \psi_{2}(\kk')\end{array}\right)
    \label{eq:sym-c1-coupledodd}
\end{equation}
for the odd-parity and, similarly, for even-parity gap functions,
\begin{equation}
  \left(\begin{array}{c}\psi_{3}(\kk)\\ d_{2}^{z}(\kk)\end{array}\right) = -T\sum_{n, \kk'}4v^{+}_{\kk \kk'}\left[M(\kk')\right]\left(\begin{array}{c}\psi_{3}(\kk')\\ d_{2}^{z}(\kk')\end{array}\right).
    \label{eq:coupledeven}
\end{equation}
The matrix  in Eqs.~(\ref{eq:sym-c1-coupledodd}) and (\ref{eq:coupledeven}) is given by
\begin{eqnarray}
    M_{11}(\kk) &=& G_{0+}\tilde{G}_{0+} + G_{0-}\tilde{G}_{0-}-2 \hat{g}_k^2 G_{0-}\tilde{G}_{0-},\label{eq:m11}\\
	M_{22}(\kk) &=& G_{0+}\tilde{G}_{0+} + G_{0-}\tilde{G}_{0-}-2 \hat{\varepsilon}_{\kk}^2 G_{0-}\tilde{G}_{0-},\label{eq:m22}\\
    M_{12}(\kk) &=&  2i \hat{g}_{\kk}  \hat{\varepsilon}_{\kk}G_{0-}\tilde{G}_{0-} = M^{*}_{21}(\kk).
    \label{eq:m12}
\end{eqnarray}

Performing the sums over the Matsubara frequencies, we first discuss the uncoupled $x$ and $y$ component of the $d$ vector and choose the odd-parity gap functions, $ \vec{d}_{3}(\kk)= (\Delta^{x}_{-} \hat{x} + \Delta^{y}_{-} \hat{y} )\sin k_{x}$ obtained for a nearest-neighbor interaction [a degenerate solution of the linearized gap equation is  $ \vec{d}_{3}(\kk)= (\Delta^{x}_{-} \hat{x} + \Delta^{y}_{-} \hat{y} ) \sin k_{y}$]. Eq.~\eqref{eq:sym-c1-eo-xy2} yields then the standard BCS equation determining $ T_c $ which is degenerate for both $x$ and $y$ component, 
\begin{equation}
    1 = -V\sum_{\kk'}\sum_{a=\pm}\frac{\sin^{2}k_{x}'}{2 \xi_{a,\kk'}}\tanh\left( \frac{\xi_{a,\kk'}}{2T} \right) .
    \label{eq:sym-c1-dxym}
\end{equation}

We turn now to the even-parity gap function $ \vec{d}_2 (\kk) $ for which we assume an extended-$s$-wave form,
$\vec{d}_{2}(\kk)= (\Delta^{x}_{+} \hat{x} + \Delta^{y}_{+} \hat{y} ) (\cos k_x + \cos k_y) $, a result of the nearest-neighbor interaction. 
This leads to the 
equation for $ T_c $,
\begin{equation}
 1 = -V\sum_{\kk'}\sum_{a=\pm}\frac{(\cos k_{x}' + \cos k_{y}')^{2}}{2(\varepsilon_{\kk'}^{\rm intra}-\mu)}\tanh\left( \frac{\xi_{a,\kk'}}{2T} \right) ,
    \label{eq:sym-c1-dxyp}
\end{equation}
originating from Eq.~(\ref{eq:sym-c1-eo-xy1}).

These equations should be compared with the corresponding equations (\ref{eq:sym-c1-coupledodd}) and (\ref{eq:coupledeven}). For the odd-parity case with the nearest-neighbor coupling approach and
the gap functions $d_{3}^{z}(\kk)=\Delta^{z}_{-}\sin k_{x}$ and $\psi_{2}(\kk)=\Delta_{-}^{s}\sin k_{x}$, we obtain from Eq.~(\ref{eq:sym-c1-coupledodd}) 
\begin{equation}
    \left(\begin{array}{c}\Delta^{z}_{-}\\ \Delta_{-}^{s}\end{array}\right) = \left(\begin{array}{cc}L^{-}_{0} + L^{-}_{1}& iL^{-}_{3}\\-iL^{-}_{3} & L^{-}_{0} + L^{-}_{2}\end{array}\right)\left(\begin{array}{c}\Delta_{-}^{z}\\ \Delta_{-}^{s}\end{array}\right).
    \label{eq:sym-c1-coupledeven}
\end{equation}
Summing again over the Matsubara frequencies, we can express these matrix elements as
\begin{eqnarray}
    L^{-}_{0} &=& -V\sum_{\kk}\sin^{2} k_{x} S_{1}(\kk),\nonumber\\
    L^{-}_{1} &=& -V\sum_{\kk}\sin^{2} k_{x} \hat{g}^2_{\kk}[S_{2}(\kk) - S_{1}(\kk)],\nonumber\\
    L^{-}_{2} &=& -V\sum_{\kk}\sin^{2} k_{x} \hat{\varepsilon}_{\kk}^{2}[S_{2}(\kk) - S_{1}(\kk)],\nonumber\\
    L^{-}_{3} &=& -V\sum_{\kk}\sin^{2} k_{x} \hat{g}_{\kk}\hat{\varepsilon}_{\kk}[S_{2}(\kk) - S_{1}(\kk)]
    \label{eqn:sym-c1-coupleelements}
\end{eqnarray}
with
\begin{eqnarray}
    S_{1}(\kk) &=& \sum_{a=\pm}\frac{1}{2\xi_{a, \kk}}\tanh\left( \frac{\xi_{a,\kk}}{2T} \right)\\
    S_{2}(\kk) &=& \sum_{a=\pm}\frac{1}{2(\varepsilon_{\kk}^{\rm intra}-\mu)}\tanh\left( \frac{\xi_{a,\kk}}{2T} \right).
    \label{eq:mats1}
\end{eqnarray}
The analogous result can be obtained for the even-parity case with
$ d_2^z (\kk) = \Delta^z_+ (\cos k_x + \cos k_y) $ and $ \psi_3(\kk) = \Delta^s_+ (\cos k_x + \cos k_y) $, leading to
\begin{equation}
    \left(\begin{array}{c}\Delta_{+}^{s}\\\Delta^{z}_{+}\end{array}\right) = \left(\begin{array}{cc}L^{+}_{0} + L^{+}_{1}& iL^{+}_{3}\\-iL^{+}_{3} & L^{+}_{0} + L^{+}_{2}\end{array}\right)\left(\begin{array}{c}\Delta_{+}^{s}\\\Delta^{z}_{+}\end{array}\right)
    \label{eq:sym-c2-coupledodd}
\end{equation}
with
\begin{eqnarray}
    L^{+}_{0} \!\!&=&\!\! -V\!\sum_{\kk}(\cos k_x\! +\! \cos k_y)^2 S_{1}(\kk),\nonumber\\
    L^{+}_{1} \!\!&=&\!\! -V\!\sum_{\kk}(\cos k_x\! +\! \cos k_y)^2 \hat{g}^2_{\kk}[S_{2}(\kk)\! -\! S_{1}(\kk)],\nonumber\\
    L^{+}_{2} \!\!&=&\!\! -V\!\sum_{\kk}(\cos k_x\! +\! \cos k_y)^2 \hat{\varepsilon}_{\kk}^{2}[S_{2}(\kk)\! -\! S_{1}(\kk)],\nonumber\\
    L^{+}_{3} \!\!&=&\!\! -V\!\sum_{\kk}(\cos k_x\! +\! \cos k_y)^2 \hat{g}_{\kk}\hat{\varepsilon}_{\kk}[S_{2}(\kk)\! -\! S_{1}(\kk)] .
    \label{eqn:sym-c2-coupleelements}
\end{eqnarray}
The instability condition for Eq.~(\ref{eq:sym-c1-coupledeven}) and (\ref{eq:sym-c2-coupledodd}) are given by the eigenvalues 
\begin{equation}
\lambda_{\pm}^s = L_0^s + \frac{1}{2} (L_1^s + L_2^s ) \pm \frac{1}{2} \sqrt{ (L_1^s  - L_2^s )^2 + (2L_3^s)^2 }
\end{equation}
reaching $ \lambda_{\pm}^s = 1 $ for both even and odd parity with $ s = + $ [Eq.~(\ref{eq:sym-c1-coupledeven})]
and $ s= - $ [Eq.~(\ref{eq:sym-c2-coupledodd})]. 
Note that the instability condition for the $x$ and $y$ components in Eq.~(\ref{eq:sym-c1-dxym}) and (\ref{eq:sym-c1-dxyp}) correspond to
\begin{equation}
\lambda_1 = L_0^- = 1
\end{equation}
for odd-parity pairing [Eq.~(\ref{eq:sym-c1-dxym})] and
\begin{equation}
\lambda_2 = L_0^+ + L_1^+ + L_2^+ = 1
\end{equation}
for even-parity pairing [Eq.~(\ref{eq:sym-c1-dxyp})]. 
It can be demonstrated easily that $ L_1^s , L_2^s < 0 $ and $ L_1^s L_2^s \geq (L_3^{s})^2 $.  We now use the resulting inequality, 
\begin{equation}
(L_1^s-L_2^s)^2 + (2L_3^s)^2 \leq (L_1^s + L_2^s)^2
\end{equation}
and $ \lambda_+^s \geq \lambda_-^s $ to obtain the relation:
\begin{equation}
L_0^s \geq \lambda_+^s \geq \lambda_-^s \geq L_0^s + L_1^s + L_2^s .
\end{equation}
 From these relations we are able to show for the odd-parity states
\begin{equation}
\lambda_1^- = L_0^- \geq \lambda_{\pm}^-
\end{equation}
such that the instability leads to a state with the $d$ vector perpendicular to the $z$ axis as described by
Eq.~(\ref{eq:sym-c1-eo-xy2}). On the other hand, for even-parity pairing the inequality,
\begin{equation}
\lambda_{\pm}^+ \geq \lambda_2^- = L_0^s + L_1^s + L_2^s
\end{equation}
favors the state with spin-singlet and -triplet mixing where the $d$ vector points in $z$ direction as described by Eq.~(\ref{eq:sym-c2-coupledodd}).

\section{Staggered non-centrosymmetric plaquette structures}
\label{sec:feas}
\begin{figure}[tb]
    \begin{center}
	\includegraphics{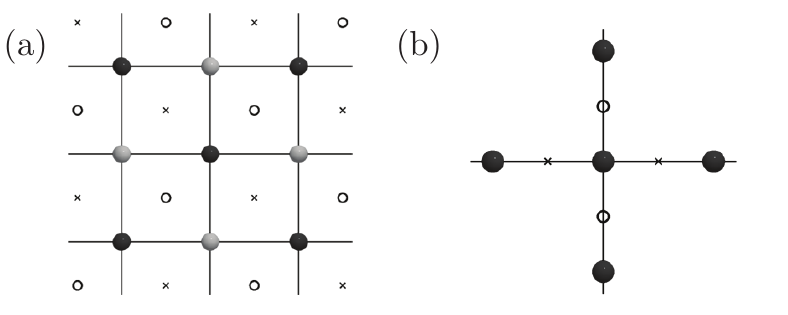}
    \end{center}
    \caption{(a) Top view of the basic FeAs crystal structure. The open circles denote As ions lying below the plane while the crosses denote ions above the plane. (b) Top view of one of the sublattices rotated by 45 degrees for an easier analysis of the hopping Hamiltonian.}
    \label{fig:feas}
\end{figure}
As a further application we turn to a system with two sublattices each lacking inversion symmetry, motivated by the crystal structure of some of the iron-pnictide superconductors. There, a single FeAs layer consists of Fe ions forming a square lattice with As ions sitting in every center of the squares. As is depicted in Fig.~\ref{fig:feas}(a), the As are shifted out of the Fe plane in a way as to built distorted tetrahedral cages around the Fe sites. Due to the arrangement of the As sites, this structure can again be described with two sublattices of checker-board type. 

\subsection{Analysis of symmetry}
As in the first example of inversion-symmetry-lacking layers, this crystal possesses a center of inversion that is located between the sublattices. Here, however, the crystal structure is non-symmorphic. Taking the symmetry center on one of the Fe sites, we can again separate the symmetry operations within $D_{4h}$ leaving the sublattice structure invariant and the ones interchanging the sublattices,
\begin{equation} \begin{array}{l}
G^{\rm intra} = \{ E, C_2, C_2' ,2 S_4 , 2 \sigma_d \} = D_{2d} , \\ \\
G^{\rm inter} = \{2 C_4, 2C_2'', I, \sigma_h, 2 \sigma_v,\} .
\end{array} \end{equation}
As $G^{\rm inter}$ contains inversion, the representation of $D_{4h}$ that changes sign for all elements of $G^{\rm inter}$ is again odd, namely $\Gamma' = B_{1u}$.

Following the characterization of the terms in the Hamiltonian introduced above, a symmetry-reducing term has to be of $B_{1u}\otimes B_{1u}$ symmetry. Such a  term is microscopically derived in appendix \ref{app:feas} for a simplified orbital structure on the Fe sites and has the same orbital and spin structure as the one in the first example, however with
\begin{equation}
    \vec{f}_{\kk}^{\,1} = \alpha(\hat{x} \sin k_x \cos k_y - \hat{y} \cos k_x \sin k_y).
    \label{eq:c2-soc}
\end{equation}
The symmetry allowed couplings between pairing states of different symmetry can again be found factorizing the gap functions according to 
\begin{equation}
  \Gamma\otimes A_{1g} \leftrightarrow (\Gamma\times B_{1u}) \otimes B_{1u}
  \label{eq:feas-coupling}
\end{equation}
leading, for example, to
\begin{eqnarray}
    (A_{1g}\otimes A_{1g}) &\leftrightarrow& (B_{1u}\otimes B_{1u})\\
    (A_{1g}\otimes B_{1u}) &\leftrightarrow& (B_{1u}\otimes A_{1g}).
    \label{eq:couplings2}
\end{eqnarray}
This yields again states of mixed parity. 
\subsection{Microscopic considerations}
The Hamiltonian describing this system has the same structure as the one encountered in Sec.~\ref{sec:layers}, but with different dispersions given by
\begin{eqnarray}
    \varepsilon_{\kk}^{\rm intra} &=& -4t'\cos k_{x} \cos k_{y},\\
    \varepsilon_{\kk}^{\rm inter} &=& -2t(\cos k_{x} + \cos k_{y}),
    \label{eq:c2-eps}
\end{eqnarray}
and $\vec{f}_{\kk}^{\,1}$ as defined in Eq.~\eqref{eq:c2-soc}. The linearized gap equations have therefore the form of Eqs.~\eqref{eq:c1-3d-singlet1}-\eqref{eq:c1-3d-triplet} for the intra-sublattice pairing and Eqs.~\eqref{eq:c1-even-singlet} and \eqref{eq:c1-odd-triplet} for the inter-sublattice pairing.
Note that the difference in the crystal structure has more drastic consequences, since the possible pairing terms in the interaction for intra- and inter-sublattice interactions allow now for different gap functions as summarized in Table~\ref{tab:c2-gapfunctions}. Obviously, the leading pairing channel has to be at least a next-nearest-neighbor interaction to allow for spin-singlet to spin-triplet mixing for the intra-sublattice pairing. Therefore, we find for example a coupling of the two gaps
\begin{multline}
  A_{1g}: \psi_0(\kk) = \psi_0 \cos k_x \cos k_y \leftrightarrow \\
  B_{1u}:\vec{d}_1(\kk)=d_1(\hat{y}\sin k_x\cos k_y - \hat{x} \sin k_y \cos k_x).
  \label{eq:spm}
\end{multline}

Considering inter-sublattice pairing, the dominant channel is modeled by a nearest-neighbor interaction. Regarding first spin-triplet pairing, spin-orbit coupling lifts the degeneracy of the spin configuration. Obviously, states belonging to the representation $ E_u $ with $ \vec{d} \parallel \hat{z} $ are according to Eq.~(\ref{eq:c1-odd-triplet}) unaffected and would yield the highest transition temperature. Any state in the other representations ($ A_{1u}, A_{2u},B_{1u}, B_{2u} $) would have a reduced $ T_c $. 
On the other hand, the spin singlet pairing channels based on inter-sublattice interactions are generally suppressed irrespective of the representation.

\begin{table}[bt]
    \centering
\begin{tabular}{c|c|c}
  & \;\;\; intra-sublattice \;\;\;& \;\;\;inter-sublattice\;\;\;\\
	 \hline
	 \;\;$A_{1g}$ \;\;& $1$, $\cos k_x\cos k_y$ & $\cos k_x + \cos k_y$\\
	 $B_{1g}$ & - & $\cos k_x - \cos k_y$\\
	  $B_{2g}$ & $ \sin k_x \sin k_y $ & -\\ [0.5ex]
	 \hline
	 $A_{1u}$ & $\hat{x}\sin k_x \cos k_y + \hat{y}\sin k_y \cos k_x $ & $\hat{x}\sin k_x + \hat{y}\sin k_y$ \\
	 $A_{2u}$ & $\hat{y}\sin k_x \cos k_y - \hat{x}\sin k_y \cos k_x $ & $\hat{x}\sin k_y - \hat{y}\sin k_x$ \\
	 $B_{1u}$ & $\hat{x}\sin k_x \cos k_y - \hat{y}\sin k_y \cos k_x $ & $\hat{x}\sin k_x - \hat{y}\sin k_y$ \\
	 $B_{2u}$ & $\hat{y}\sin k_x \cos k_y + \hat{x}\sin k_y \cos k_x $ & $\hat{x}\sin k_y + \hat{y}\sin k_x$ \\
	  $ E_u $ & $ \{\hat{z} \sin k_x \cos k_y , \hat{z} \sin k_y \cos k_x \} $ & $ \{ \hat{z} \sin k_x , \hat{z} \sin k_y \} $
 \end{tabular}
\caption{Lowest order basis functions supported by intra- and inter-sublattice interactions on the lattice considered in Sec.~\ref{sec:feas}, i.e., for on-site and nearest-neighbor interactions, respectively. In order to allow for a spin-singlet to spin-triplet coupling, an interaction between next-to-nearest neighbors has to be considered.}
    \label{tab:c2-gapfunctions}
\end{table}

\section{Conclusion}
For crystal lattices, where inversion symmetry is broken in a regular, but non-uniform (unit-cell multiplying) pattern, the multi-band structure of the reduced Brillouin zone renders the classification of the superconducting order parameter in terms of standard spin-singlet and spin-triplet insufficient. On the level of the  normal state electronic properties this is imprinted by
spin-orbit coupling whose structure is closely connected to lattice-symmetry details. 
For such systems, the usual connection between even (odd) parity in momentum space and spin-singlet (spin-triplet) configuration is lost in many cases, although the overall system is centrosymmetric. For a full classification in terms of the crystal symmetry, also the structure of the gap in band space has to be taken into account.

In this paper, we have studied two main classes of a local lack of inversion symmetry with a two-sublattice structure, whereby the complete lattice possesses a center of inversion. This corresponds to a doubling of the unit cell leading to a two-band description. The first class of lattices is characterized by the
property that each sublattice has broken inversion symmetry. This manifests itself in the symmetry group $ G^{\rm intra} $ which does not include the element of inversion $ I $. This is the case for our first (Sect.\ref{sec:layers}) and last (Sect.\ref{sec:feas}) example. The situation yields singlet-triplet mixing which is characterized by the lattice specific representation $ \Gamma' $ having odd parity and occurs in the connection with intra-sublattice pairing. 

In the other 
case, the sublattices retain separately inversion symmetry, i.e. $ G^{\rm intra} $ contains $ I $, while the links connecting the sublattices lack inversion symmetry. The corresponding representation $ \Gamma' $ has even parity,
which also determines the structure of the spin-orbit coupling. In this system, it is inter-sublattice pairing which 
mixes spin-singlet and spin-triplet pairing while the parity remains fixed. Here obviously spin-configuration and parity are not
anymore tied together. 

This new classification scheme can be important to determine which pairing states can be stabilized. This can be particularly useful, if questions concerning the degeneracy in spin space have to be answered. The new states and electronic structures may have an impact on the way superconductors couple (Josephson effect) and how 
the superconducting state reacts on external magnetic fields. These topics will be discussed elsewhere.

\section*{Acknowledgements}
We are grateful for many helpful discussions to D.F. Agterberg, D. Maruyama, and Y. Yanase. This work was financially supported by the Swiss Nationalfonds and the NCCR MaNEP. M.H.F. acknowledges support from the NSF Grant DMR-0520404 to the Cornell Center for Materials Research and NSF Grant DMR-0955822 and F.L. acknowledges support from the DFG through TRR 80.

\appendix
\section{Structure of Interaction}
\label{app:interaction}
In this appendix, the structure of a general density-density interaction in a crystal with a two-site unit cell is analyzed. The generalization to other types of interactions, e.g., a spin-spin interaction, is straightforward. Our starting point is a real-space formulation of the interaction,
\begin{equation}
    \HH' = \sum_{i, j}\sum_{s,s'}V_{ij}n_{i s}n_{j s'}=\sum_{i, j}\sum_{s,s'}V_{ij}c_{i s}^{\dag}c_{j s'}^{\dag}c_{j s'}^{\phantom{\dag}}c_{i s}^{\phantom{\dag}}
    \label{eq:general-int}
\end{equation}
with $V_{ij}$ the interaction strength between the lattice sites $i$ and $j$. Note that for the special case of $i=j$ (on-site interaction) the spin sum only runs over $s\neq s'$.

Changing to momentum space, Eq.~(\ref{eq:general-int}) yields
\begin{equation}
    \HH' = \frac{1}{N}\sum_{\kk,\kk', \q}\sum_{s, s'}v(\kk,\kk')c_{\kk s}^{\dag}c_{-\kk+\q s'}^{\dag}c_{-\kk'+\q s'}^{\phantom{\dag}}c_{\kk's}^{\phantom{\dag}},
    \label{eq:momentum-int}
\end{equation}
where $v(\kk,\kk')=v(\kk - \kk')$ due to translational symmetry of the crystal. Since we are interested in a situation with two sites per unit cell, we introduce two species of electron operators,
\begin{equation}
    c_{\alpha \kk s}^{\dag} = \left\{\begin{array}{ll}c_{\kk s}^{\dag} & \alpha =1, \\ c_{\kk+\Q s}^{\dag}&\alpha=2,\end{array}\right.
    \label{eq:species}
\end{equation}
where $\Q = (0,0,\pi)$ for a system as described in section~\ref{sec:c1} and $\Q=(\pi,\pi)$ for the situation of sections~\ref{sec:bonds} and \ref{sec:feas}, respectively. Accordingly, we restrict the sum in Eq.~(\ref{eq:momentum-int}) to the two cases of $\q={\textbf 0}$ and $\q= \Q$ in the following. For the case $\q={\textbf 0}$ we find
\begin{multline}
    \HH_{\bf 0}' = \frac{1}{N}\sum_{\kk\kk'}\{v(\kk-\kk')[c_{1\kk s}^{\dag}c_{1-\kk s'}^{\dag}c_{1-\kk's'}^{\phantom{\dag}}c_{1\kk's}^{\phantom{\dag}}\\ + c_{2\kk s}^{\dag}c_{2-\kk s'}^{\dag}c_{2-\kk' s'}^{\phantom{\dag}}c_{2\kk's}^{\phantom{\dag}}]\\
    + v(\kk-\kk'+\Q)[c_{1\kk s}^{\dag}c_{1-\kk s'}^{\dag}c_{2-\kk's'}^{\phantom{\dag}}c_{2\kk's}^{\phantom{\dag}}\\ + c_{2\kk s}^{\dag}c_{2-\kk s'}^{\dag}c_{1-\kk's'}^{\phantom{\dag}}c_{1\kk's}^{\phantom{\dag}}]\}.
    \label{eq:int_q0}
\end{multline}
For the other case $\q=\Q$, the interaction term can similarly be written as
\begin{multline}
    \HH_{\Q}' = \frac{1}{N}\sum_{\kk\kk'}\{v(\kk-\kk')[c_{1\kk s}^{\dag}c_{2-\kk s'}^{\dag}c_{2-\kk's'}^{\phantom{\dag}}c_{1\kk's}^{\phantom{\dag}}\\
    + c_{2\kk s}^{\dag}c_{1-\kk s'}^{\dag}c_{1-\kk' s'}^{\phantom{\dag}}c_{2\kk's}^{\phantom{\dag}}]\\
    + v(\kk-\kk'+\Q)[c_{1\kk s}^{\dag}c_{2-\kk s'}^{\dag}c_{1-\kk's'}^{\phantom{\dag}}c_{2\kk's}^{\phantom{\dag}}\\
    + c_{2\kk s}^{\dag}c_{1-\kk s'}^{\dag}c_{2-\kk's'}^{\phantom{\dag}}c_{1\kk's}^{\phantom{\dag}}]\}.
    \label{eq:int_qQ}
\end{multline}
At this point, we can distinguish the two cases of an interaction between sites belonging to the same sublattice and between sites on different sublattices. For the former case, $i, j \in \mathcal{A}$ ($\mathcal{B}$), we can use $v(\kk +\Q) = v(\kk)$ to write the above expressions as
\begin{equation}
  \HH_{\bf 0, Q}' = \frac{1}{N}\sum_{\kk, \kk'}v^{\bf 0, Q}_{\alpha\beta\gamma\delta}(\kk-\kk')c_{\alpha \kk s}^{\dag}c_{\beta-\kk s'}^{\dag}c_{\gamma-\kk's'}^{\phantom{\dag}}c_{\delta \kk's}^{\phantom{\dag}},\\
  \label{eq:appint0Q1>}
\end{equation}
with
\begin{eqnarray}
      v^{\bf 0}_{\alpha\beta\gamma\delta} &=& v(\kk-\kk')[(\tau^0)_{\alpha\beta}(\tau^0)^{\dag}_{\gamma\delta}],\\
      v^{\bf Q}_{\alpha\beta\gamma\delta} &=& v(\kk-\kk')[(\tau^1)_{\alpha\beta}(\tau^1)^{\dag}_{\gamma\delta}].
    \label{eq:int_qa}
    \end{eqnarray}
Similarly, for the latter case, where $i\in\mathcal{A}(\mathcal{B})$ and $j \in \mathcal{B} (\mathcal{A})$, $v(\kk +\Q) = -v(\kk)$ yields
\begin{eqnarray}
	v^{\bf 0}_{\alpha\beta\gamma\delta}&=& v(\kk-\kk')[(\tau^3)_{\alpha\beta}(\tau^3)^{\dag}_{\gamma\delta}],\\
	v^{\bf Q}_{\alpha\beta\gamma\delta} &=& v(\kk-\kk')[(i\tau^2)_{\alpha\beta}(i\tau^2)^{\dag}_{\gamma\delta}].
        \label{eq:int_qb}
    \end{eqnarray}
In addition, the interaction can also be separated in a spin-singlet and a spin-triplet channel introducing Pauli matrices for the spin degrees of freedom,
\begin{multline}
    \sum_{ss'}c_{\alpha \kk s}^{\dag}c_{\beta-\kk s'}^{\dag}c_{\gamma-\kk's'}^{\phantom{\dag}}c_{\delta \kk's}^{\phantom{\dag}}\\
    =\frac{1}{2} \sum_{s_1\dots s_4}\Lambda_{s_1s_2s_3s_4}c_{\alpha \kk s_1}^{\dag}c_{\beta-\kk s_2}^{\dag}c_{\gamma-\kk's_3}^{\phantom{\dag}}c_{\delta \kk's_4}^{\phantom{\dag}},
    \label{eq:spins}
\end{multline}
where
\begin{equation}
    \Lambda_{s_1s_2s_3s_4} = (\varsigma^0)_{s_1s_2}(\varsigma^0)^{\dag}_{s_3s_4} + (\vec{\varsigma})_{s_1s_2}\cdot(\vec{\varsigma})^{\dag}_{s_3s_4}.
    \label{eq:spin-part}
\end{equation}
Here, we have introduced $\varsigma^{0} = i \sigma^{y}$ and $\vec{\varsigma} = \vec{\sigma}i\sigma^{y}$ for simplicity of notation.

The total interaction term now has the form
\begin{equation}
	\HH'=\frac{1}{N}\sum_{\kk, \kk'}[V(\kk, \kk')]_{\alpha\beta\gamma\delta}^{s_1s_2s_3s_4}c_{\alpha \kk s_1}^{\dag}c_{\beta-\kk s_2}^{\dag}c_{\gamma-\kk's_3}^{\phantom{\dag}}c_{\delta \kk's_4}^{\phantom{\dag}}.
    \label{eq:int-tot}
\end{equation}
The interaction matrix element $[V(\kk, \kk')]_{\alpha\beta\gamma\delta}^{s_1s_2s_3s_4}$ has an odd and an even part in $\kk$ which depends on the resulting sign of an interchange of the two first index pairs, $(\alpha\beta, s_1s_2) \leftrightarrow(\beta\alpha, s_2s_1)$,
\begin{multline}
  [V(\kk, \kk')]_{\alpha\beta\gamma\delta}^{s_1s_2s_3s_4} =v^{+}_{\kk \kk'}\Lambda_{+,\alpha\beta\gamma\delta}^{s_1s_2s_3s_4}\\
  + v^{-}_{\kk \kk'}\Lambda_{-,\alpha\beta\gamma\delta}^{s_1s_2s_3s_4},
    \label{eq:int-split}
\end{multline}
where 
\begin{equation}
  v^{\pm}_{\kk,\kk'} = \frac{1}{2}(v(\kk - \kk') \pm v(\kk + \kk')).\nonumber\\
    \label{eq:interplus}
\end{equation}
For the intra-sublattice interaction, these read
\begin{multline}
    \Lambda_{+,\alpha\beta\gamma\delta}^{s_1s_2s_3s_4}=(\varsigma^0)_{s_1s_2}(\varsigma^0)^{\dag}_{s_3s_4}\times\\
    \times [(\tau^0)_{\alpha\beta}(\tau^0)^{\dag}_{\gamma\delta}+(\tau^1)_{\alpha\beta}(\tau^1)^{\dag}_{\gamma\delta}]
    \label{eq:intra-plus}
\end{multline}
and 
\begin{multline}
  \Lambda_{-,\alpha\beta\gamma\delta}^{s_1s_2s_3s_4}=(\vec{\varsigma})_{s_1s_2}\cdot(\vec{\varsigma})^{\dag}_{s_3s_4}\times\\
    \times[(\tau^0)_{\alpha\beta}(\tau^0)^{\dag}_{\gamma\delta} + (\tau^1)_{\alpha\beta}(\tau^1)^{\dag}_{\gamma\delta}],
    \label{eq:intra-minus}
\end{multline}
while for the inter-sublattice interaction, we find
\begin{multline}
    \Lambda_{+,\alpha\beta\gamma\delta}^{s_1s_2s_3s_4}=[(\tau^3)_{\alpha\beta}(\tau^3)^{\dag}_{\gamma\delta}](\varsigma^0)_{s_1s_2}(\varsigma^0)^{\dag}_{s_3s_4}\\
    + [(i \tau^2)_{\alpha\beta}(i\tau^2)^{\dag}_{\gamma\delta}](\vec{\varsigma})_{s_1s_2}\cdot(\vec{\varsigma})^{\dag}_{s_3s_4}
    \label{eq:inter-plus}
\end{multline}
and 
\begin{multline}
    \Lambda_{-,\alpha\beta\gamma\delta}^{s_1s_2s_3s_4}=[(i \tau^2)_{\alpha\beta}(i\tau^2)^{\dag}_{\gamma\delta}](\varsigma^0)_{s_1s_2}(\varsigma^0)^{\dag}_{s_3s_4}\\
    +[(\tau^3)_{\alpha\beta}(\tau^3)^{\dag}_{\gamma\delta}](\vec{\varsigma})_{s_1s_2}\cdot(\vec{\varsigma})^{\dag}_{s_3s_4},
    \label{eq:inter-minus}
\end{multline}
Unlike the case of a primitive unit cell, the momentum dependence is thus not only depending on the spin part of the interaction.

As an example, we look in the following at the specific example of stacked layers of Sec.~\ref{sec:c1}. The simplest non-trivial intra-sublattice interaction is between nearest neighbors, i.e., $v(\kk - \kk') = V[\cos(k_x-k_x')+\cos(k_y - k_y')]=-v(\kk -\kk' +\Q)$.
\begin{multline}
  v^{+}_{\kk\kk'} = \frac{V}{2}(\cos k_x\! +\! \cos k_y)(\cos k_x'\! +\! \cos k_y')\\
    + \frac V2 (\cos k_x\! -\! \cos k_y')(\cos k_x'\! -\! \cos k_y')
    \label{eq:plus}
\end{multline}
and 
\begin{equation}
  v^{-}_{\kk,\kk'} = -V(\sin k_x \sin k_x' + \sin k_y \sin k_y').
    \label{int-minus}
\end{equation}
Note that for the cases of Secs.~\ref{sec:bonds} and \ref{sec:feas}, the above functions correspond to the inter-sublattice interaction.

For the nearest-neighbor inter-sublattice interaction, we find
\begin{equation}
  v^{+}_{\kk \kk'} = V \cos k_z \cos k_z'
  \label{eq:internnp}
\end{equation}
and
\begin{equation}
  v^{-}_{\kk \kk'} = -V\sin k_z \sin k_{z}'.
  \label{eq:internnm}
\end{equation}

\section{Hopping matrix elements in systems like Fe-As-compounds}
\label{app:feas}
The special structure of the FeAs layers in the iron-pnictides leads to a spin-orbit coupling with a different sign depending on the sublattice. In this appendix, this spin-orbit coupling is derived for a simplified orbital structure, considering $s$-like orbitals for the Fe sites and $p$-type orbitals for the As ions, by focussing on only one sublattice [see Fig.~\ref{fig:feas}(a)].
To analyze the nearest-neighbor hopping - corresponding to a next-nearest-neighbor hopping in the full structure - it is easiest to rotate the crystal by 45 degrees and start with the As ions first lying on the bonds [see Fig.~\ref{fig:feas}(b)]. For this situation, the electrons can only hop from one Fe to the next in $x$ ($y$) direction via a $p_{x}$ ($p_{y}$) orbital with hopping element $t_{sp}$, 
\begin{multline}
    \HH_{\rm nnn} = -t_{sp} \sum_{i, s}[c^{\dag}_{i s}p^{(x)}_{i + \hat{x}/2 s} - c^{\dag}_{i s}p^{(x)}_{i - \hat{x}/2 s}\\
    + c^{\dag}_{i s}p^{(y)}_{i + \hat{y}/2 s} - c^{\dag}_{i s}p^{(y)}_{i - \hat{y}/2 s}) + {\rm h.c.}].
    \label{eq:nnnhop0}
\end{multline}

Assuming that the As-orbital's on-site energy differs from the energy of the Fe orbitals, $E_{\rm As} = E_{\rm Fe} - \Delta$, we find for the nearest-neighbor-hopping integral in the effective one-band model
\begin{equation}
    t' = \frac{t_{sp}^2}{\Delta}.
    \label{eq:nnnhopt0}
\end{equation}
The Hamiltonian in momentum space thus reads
\begin{equation}
    \HH^{\rm hop} = \sum_{\kk'}\varepsilon_{\kk'}^{\rm hop}c_{\kk' s}^{\dag}c_{\kk' s}^{\phantom{\dag}},
    \label{eq:nnnhopmom0}
\end{equation}
where $\varepsilon_{\kk'}^{\rm hop} = -2t'(\cos k_x' + \cos k_y')$ with the new rotated axes $k_x'$ and $k_y'$. Rotating the crystal back by 45 degrees to change to the old axes we find using $k_x' = (k_x - k_y)$, $k_y' = (k_x + k_y)$ and
\begin{equation}
    \cos(k_x \pm k_y) = \cos k_x \cos k_y \mp \sin k_x \sin k_y
    \label{eq:cospm}
\end{equation}
the usual (nnn) hopping energy $\varepsilon_{\kk}^{\rm hop} = -4t'\cos k_x \cos k_y$.

If the As ions are moved out of the plane, it becomes also possible to hop via a $p_{z}$ to a neighboring Fe site with hopping integral $\tilde{t}_{sp}$. We therefore find the additional hoppings
\begin{multline}
    \HH = -\tilde{t}_{sp} \sum_{i, s}[c^{\dag}_{i s}p^{(z)}_{i + \hat{x}/2 s} + c^{\dag}_{i s}p^{(z)}_{i - \hat{x}/2 s}\\
    - c^{\dag}_{i s}p^{(z)}_{i + \hat{y}/2 s} - c^{\dag}_{i s}p^{(z)}_{i - \hat{y}/2 s}) + {\rm h.c.}].
    \label{eq:nnnhop}
\end{multline}
We can now change to eigenfunctions of the As-site SOC $p^{(\pm)}_{j s}$, where the spin-quantization axis has to be orthogonal to the hopping direction to find
\begin{multline}
    \HH = -\sum_{i s}\Big(\tilde{t}c^{\dag}_{i s}p^{(+)}_{i + \hat{x}/2 s} + \tilde{t}^{*}c^{\dag}_{i s}p^{(-)}_{i + \hat{x}/2 s}\\
    -\tilde{t}^{*}c^{\dag}_{i s} p^{(+)}_{i - \hat{x}/2 s} -\tilde{t}c_{i s}^{\dag} p^{(-)}_{i - \hat{x}/2 s}\\
    -(i\tilde{t})c^{\dag}_{i s}p^{(+)}_{i + \hat{y}/2 s} - (i\tilde{t})^{*}c^{\dag}_{i s}p^{(-)}_{i + \hat{y}/2 s}\\
    +(i\tilde{t})^{*}c^{\dag}_{i s} p^{(+)}_{i - \hat{y}/2 s} +(i\tilde{t})c_{i s}^{\dag} p^{(-)}_{i - \hat{y}/2 s}) + {\rm h.c.}\Big)\\
    \label{eq:3bhoppings}
\end{multline}
with $\tilde{t} = (t_{sp} + i \tilde{t}_{sp})/\sqrt{2}$.

Again reducing this to a single-band model by integrating out the As orbitals, we find in addition to the hopping Hamiltonian
\begin{equation}
    -t' \sum_{<i,j>}\sum_{s}(c^{\dag}_{i s}c^{\phantom{\dag}}_{j s} + {\rm h.c.})
    \label{eq:nnnhoppings}
\end{equation}
with
\begin{equation}
    t' = (t_{sp}^2 - \tilde{t}_{sp}^2)\frac{\Delta}{\Delta^2 - \lambda^2}
    \label{eq:nnnhopren}
\end{equation}
a new SOC term,
\begin{multline}
    \HH' = \sum_{i s s'}\Big(i\alpha c^{\dag}_{i s}\sigma_{s s'}^yc^{\phantom{\dag}}_{i + \hat{x} s'} - i\alpha c^{\dag}_{i s}\sigma_{s s'}^yc^{\phantom{\dag}}_{i - \hat{x} s'}\\
    +i\alpha c^{\dag}_{i s}\sigma_{s s'}^xc^{\phantom{\dag}}_{i + \hat{y} s'} - i\alpha c^{\dag}_{i s}\sigma_{s s'}^xc^{\phantom{\dag}}_{i - \hat{y} s'}  + {\rm h.c.}\Big)
    \label{eq:addhopping}
\end{multline}
with 
\begin{equation}
    \tilde{\alpha} = \frac{2t_{sp}\tilde{t}_{sp}\lambda}{\Delta^2 - \lambda^2}.
    \label{eq:alpha}
\end{equation}
In momentum space, this additional term reads
\begin{equation}
    \HH' = \sum_{\kk, s, s'}\Big(\vec{\Lambda}_{\kk}\cdot\vec{\sigma}_{ss'}\Big) c^{\dag}_{\kk s}c^{\phantom{\dag}}_{\kk s'},
    \label{eq:sochopmom}
\end{equation}
where $\vec{\Lambda}_{\kk} = 2\tilde{\alpha}(\hat{x}\sin k_{y} - \hat{y}\sin k_x)$.

To transform this back, we use
\begin{equation}
    \sin(k_x \pm k_y) = \sin k_x \cos k_y \pm \cos k_x \sin k_y
    \label{eq:trig2}
\end{equation}
and also the rotated Pauli matrices,
\begin{eqnarray}
    \sigma^x&\mapsto&\frac{\sqrt{2}}{2}(\sigma^{x} - \sigma^{y}), \\
    \sigma^y&\mapsto&\frac{\sqrt{2}}{2}(\sigma^{x} + \sigma^{y}).
    \label{eq:paulirotation}
\end{eqnarray}
Finally, we find the SOC Hamiltonian
\begin{equation}
    \HH_{\rm soc} = \sum_{\kk, s, s'}\Big(\vec{\Lambda}_{\kk}\cdot\vec{\sigma}_{ss'}\Big) c^{\dag}_{\kk s}c^{\phantom{\dag}}_{\kk s'},
    \label{eq:finalsoc}
\end{equation}
where now $\vec{\Lambda}_{\kk} = \alpha(\hat{x}\sin k_x\cos k_y -\hat{y}\cos k_x\sin k_y)$. In a crystal with $D_{4h}$ symmetry, this term belongs to the irreducible representation $B_{1u}$. 


\end{document}